\documentclass[longauth,longauthfont]{aa}  
\usepackage{adjustbox}
\usepackage{color}
\usepackage{graphicx}
\usepackage{amsmath}
\usepackage{amssymb}
\usepackage{breqn}
\usepackage{gensymb}
\usepackage{placeins}
\usepackage{natbib}
\usepackage{upgreek}
\usepackage{subfig}
\usepackage{multirow}
\usepackage{booktabs}
\usepackage{tabularx}
\usepackage{colortbl}
\usepackage{csvsimple}
\usepackage{float}
\usepackage[colorlinks=true,
linkcolor=blue,
citecolor=blue]{hyperref}
\usepackage{txfonts}
\usepackage{comment}
\usepackage{xcolor}
\usepackage{placeins}
\usepackage{multicol}
\usepackage{soul}
\usepackage[switch]{lineno}
\usepackage{textcomp}
\usepackage{tabularx}
\usepackage{tabto}
\usepackage{mathtools}
\usepackage{multirow}
\usepackage{xspace}
\usepackage[encapsulated]{CJK}
\usepackage{ucs}
\usepackage{orcidlink}
\newcommand\sbullet[1][.5]{\mathbin{\vcenter{\hbox{\scalebox{#1}{$\bullet$}}}}}

\newcommand{\tb}{$T_{\rm b}$ }

\newcommand{\rev}[1]{\textcolor{black}{#1}}
\newcommand{\jref}[1]{\textcolor{black}{#1}} 

\begin{document} 

\title{A multi-frequency study of sub-parsec jets with\\the Event Horizon Telescope}
\titlerunning{Studying sub-parsec jets with the Event Horizon Telescope}
\authorrunning{R\"oder, Wielgus et. al.}


\author{\tiny%
Jan Röder~\orcidlink{0000-0002-2426-927X}\inst{1,2} \and
Maciek Wielgus~\orcidlink{0000-0002-8635-4242}\inst{2,1} \and
Andrei P. Lobanov~\orcidlink{0000-0003-1622-1484}\inst{1} \and
Thomas P. Krichbaum~\orcidlink{0000-0002-4892-9586}\inst{1} \and
Dhanya G. Nair~\orcidlink{0000-0001-5357-7805}\inst{3,1} \and
Sang-Sung Lee~\orcidlink{0000-0002-6269-594X}\inst{4} \and
Eduardo Ros~\orcidlink{0000-0001-9503-4892}\inst{1} \and
Vincent L. Fish~\orcidlink{0000-0002-7128-9345}\inst{5} \and
Lindy Blackburn~\orcidlink{0000-0002-9030-642X}\inst{6,7} \and
Chi-kwan Chan~\orcidlink{0000-0001-6337-6126}\inst{8,9,10} \and
Sara Issaoun~\orcidlink{0000-0002-5297-921X}\inst{7,11} \and
Michael Janssen~\orcidlink{0000-0001-8685-6544}\inst{12,1} \and
Michael D. Johnson~\orcidlink{0000-0002-4120-3029}\inst{6,7} \and
Sheperd S. Doeleman~\orcidlink{0000-0002-9031-0904}\inst{6,7} \and
Geoffrey C. Bower~\orcidlink{0000-0003-4056-9982}\inst{13,14} \and
Geoffrey B. Crew~\orcidlink{0000-0002-2079-3189}\inst{5} \and
Remo P. J. Tilanus~\orcidlink{0000-0002-6514-553X}\inst{8,12,15,16} \and
Tuomas Savolainen~\orcidlink{0000-0001-6214-1085}\inst{17,18,1} \and
C. M. Violette Impellizzeri~\orcidlink{0000-0002-3443-2472}\inst{15,19} \and
Antxon Alberdi~\orcidlink{0000-0002-9371-1033}\inst{2} \and
Anne-Kathrin Baczko~\orcidlink{0000-0003-3090-3975}\inst{20,1} \and
José L. Gómez~\orcidlink{0000-0003-4190-7613}\inst{2} \and
Ru-Sen Lu~\orcidlink{0000-0002-7692-7967}\inst{21,22,1} \and
Georgios F. Paraschos~\orcidlink{0000-0001-6757-3098}\inst{1} \and
Efthalia Traianou~\orcidlink{0000-0002-1209-6500}\inst{2,1} \and
Ciriaco Goddi~\orcidlink{0000-0002-2542-7743}\inst{23,24,25,26} \and
Daewon Kim~\orcidlink{0000-0003-4997-2153}\inst{1} \and
Mikhail Lisakov~\orcidlink{0000-0001-6088-3819}\inst{27} \and
Yuri Y. Kovalev~\orcidlink{0000-0001-9303-3263}\inst{1,6} \and
Petr A. Voitsik~\orcidlink{0000-0002-1290-1629}\inst{28} \and
Kirill V. Sokolovsky~\orcidlink{0000-0001-5991-6863}\inst{29}
\centerline{---------}
Kazunori Akiyama~\orcidlink{0000-0002-9475-4254}\inst{5,30,6} \and
Ezequiel Albentosa-Ruíz~\orcidlink{0000-0002-7816-6401}\inst{31} \and
Walter Alef\inst{1} \and
Juan Carlos Algaba~\orcidlink{0000-0001-6993-1696}\inst{32} \and
Richard Anantua~\orcidlink{0000-0003-3457-7660}\inst{6,7,33} \and
Keiichi Asada~\orcidlink{0000-0001-6988-8763}\inst{34} \and
Rebecca Azulay~\orcidlink{0000-0002-2200-5393}\inst{31,35,1} \and
Uwe Bach~\orcidlink{0000-0002-7722-8412}\inst{1} \and
David Ball~\orcidlink{0009-0009-2348-6879}\inst{8} \and
Mislav Baloković~\orcidlink{0000-0003-0476-6647}\inst{36} \and
Bidisha Bandyopadhyay~\orcidlink{0000-0002-2138-8564}\inst{3} \and
John Barrett~\orcidlink{0000-0002-9290-0764}\inst{5} \and
Michi Bauböck~\orcidlink{0000-0002-5518-2812}\inst{37} \and
Bradford A. Benson~\orcidlink{0000-0002-5108-6823}\inst{38,39} \and
Dan Bintley\inst{40,41} \and
Raymond Blundell~\orcidlink{0000-0002-5929-5857}\inst{7} \and
Katherine L. Bouman~\orcidlink{0000-0003-0077-4367}\inst{42} \and
Michael Bremer\inst{43} \and
Christiaan D. Brinkerink~\orcidlink{0000-0002-2322-0749}\inst{12} \and
Roger Brissenden~\orcidlink{0000-0002-2556-0894}\inst{6,7} \and
Silke Britzen~\orcidlink{0000-0001-9240-6734}\inst{1} \and
Avery E. Broderick~\orcidlink{0000-0002-3351-760X}\inst{44,45,46} \and
Dominique Broguiere~\orcidlink{0000-0001-9151-6683}\inst{43} \and
Thomas Bronzwaer~\orcidlink{0000-0003-1151-3971}\inst{12} \and
Sandra Bustamante~\orcidlink{0000-0001-6169-1894}\inst{47} \and
Do-Young Byun~\orcidlink{0000-0003-1157-4109}\inst{4,48} \and
John E. Carlstrom~\orcidlink{0000-0002-2044-7665}\inst{49,39,50,51} \and
Chiara Ceccobello~\orcidlink{0000-0002-4767-9925}\inst{20} \and
Andrew Chael~\orcidlink{0000-0003-2966-6220}\inst{52} \and
Dominic O. Chang~\orcidlink{0000-0001-9939-5257}\inst{6,7} \and
Koushik Chatterjee~\orcidlink{0000-0002-2825-3590}\inst{6,7} \and
Shami Chatterjee~\orcidlink{0000-0002-2878-1502}\inst{53} \and
Ming-Tang Chen~\orcidlink{0000-0001-6573-3318}\inst{13} \and
Yongjun Chen~\orcidlink{0000-0001-5650-6770}\inst{21,54} \and
Xiaopeng Cheng~\orcidlink{0000-0003-4407-9868}\inst{4} \and
Ilje Cho~\orcidlink{0000-0001-6083-7521}\inst{2,4,55} \and
Pierre Christian~\orcidlink{0000-0001-6820-9941}\inst{56} \and
Nicholas S. Conroy~\orcidlink{0000-0003-2886-2377}\inst{29,7} \and
John E. Conway~\orcidlink{0000-0003-2448-9181}\inst{20} \and
James M. Cordes~\orcidlink{0000-0002-4049-1882}\inst{53} \and
Thomas M. Crawford~\orcidlink{0000-0001-9000-5013}\inst{39,49} \and
Alejandro Cruz-Osorio~\orcidlink{0000-0002-3945-6342}\inst{57,58} \and
Yuzhu Cui~\orcidlink{0000-0001-6311-4345}\inst{59,60} \and
Brandon Curd~\orcidlink{0000-0002-8650-0879}\inst{33,6,7} \and
Rohan Dahale~\orcidlink{0000-0001-6982-9034}\inst{2} \and
Jordy Davelaar~\orcidlink{0000-0002-2685-2434}\inst{61,62,12} \and
Mariafelicia De Laurentis~\orcidlink{0000-0002-9945-682X}\inst{63,58,64} \and
Roger Deane~\orcidlink{0000-0003-1027-5043}\inst{65,66,67} \and
Jessica Dempsey~\orcidlink{0000-0003-1269-9667}\inst{40,41,68} \and
Gregory Desvignes~\orcidlink{0000-0003-3922-4055}\inst{1,69} \and
Jason Dexter~\orcidlink{0000-0003-3903-0373}\inst{70} \and
Vedant Dhruv~\orcidlink{0000-0001-6765-877X}\inst{37} \and
Indu K. Dihingia~\orcidlink{0000-0002-4064-0446}\inst{60} \and
Sean Taylor Dougall~\orcidlink{0000-0002-3769-1314}\inst{8} \and
Sergio A. Dzib~\orcidlink{0000-0001-6010-6200}\inst{43,1} \and
Ralph P. Eatough~\orcidlink{0000-0001-6196-4135}\inst{71,1} \and
Razieh Emami~\orcidlink{0000-0002-2791-5011}\inst{7} \and
Heino Falcke~\orcidlink{0000-0002-2526-6724}\inst{12} \and
Joseph Farah~\orcidlink{0000-0003-4914-5625}\inst{72,73} \and
Edward Fomalont~\orcidlink{0000-0002-9036-2747}\inst{19} \and
H. Alyson Ford~\orcidlink{0000-0002-9797-0972}\inst{8} \and
Marianna Foschi~\orcidlink{0000-0001-8147-4993}\inst{2} \and
Raquel Fraga-Encinas~\orcidlink{0000-0002-5222-1361}\inst{12} \and
William T. Freeman\inst{74,75} \and
Per Friberg~\orcidlink{0000-0002-8010-8454}\inst{40,41} \and
Christian M. Fromm~\orcidlink{0000-0002-1827-1656}\inst{76,58,1} \and
Antonio Fuentes~\orcidlink{0000-0002-8773-4933}\inst{2} \and
Peter Galison~\orcidlink{0000-0002-6429-3872}\inst{6,77,78} \and
Charles F. Gammie~\orcidlink{0000-0001-7451-8935}\inst{37,29,79} \and
Roberto García~\orcidlink{0000-0002-6584-7443}\inst{43} \and
Olivier Gentaz~\orcidlink{0000-0002-0115-4605}\inst{43} \and
Boris Georgiev~\orcidlink{0000-0002-3586-6424}\inst{8} \and
Roman Gold~\orcidlink{0000-0003-2492-1966}\inst{80,81,82} \and
Arturo I. Gómez-Ruiz~\orcidlink{0000-0001-9395-1670}\inst{83,84} \and
Minfeng Gu~\orcidlink{0000-0002-4455-6946}\inst{21,85} \and
Mark Gurwell~\orcidlink{0000-0003-0685-3621}\inst{7} \and
Kazuhiro Hada~\orcidlink{0000-0001-6906-772X}\inst{86,87} \and
Daryl Haggard~\orcidlink{0000-0001-6803-2138}\inst{88,89} \and
Kari Haworth\inst{7} \and
Michael H. Hecht~\orcidlink{0000-0002-4114-4583}\inst{5} \and
Ronald Hesper~\orcidlink{0000-0003-1918-6098}\inst{90} \and
Dirk Heumann~\orcidlink{0000-0002-7671-0047}\inst{8} \and
Luis C. Ho~\orcidlink{0000-0001-6947-5846}\inst{91,92} \and
Paul Ho~\orcidlink{0000-0002-3412-4306}\inst{34,41,40} \and
Mareki Honma~\orcidlink{0000-0003-4058-9000}\inst{86,87,93} \and
Chih-Wei L. Huang~\orcidlink{0000-0001-5641-3953}\inst{34} \and
Lei Huang~\orcidlink{0000-0002-1923-227X}\inst{21,85} \and
David H. Hughes\inst{83} \and
Shiro Ikeda~\orcidlink{0000-0002-2462-1448}\inst{30,94,95,96} \and
Makoto Inoue~\orcidlink{0000-0001-5037-3989}\inst{34} \and
David J. James~\orcidlink{0000-0001-5160-4486}\inst{97,98} \and
Buell T. Jannuzi~\orcidlink{0000-0002-1578-6582}\inst{8} \and
Britton Jeter~\orcidlink{0000-0003-2847-1712}\inst{34} \and
Wu Jiang~\orcidlink{0000-0001-7369-3539}\inst{21} \and
Alejandra Jiménez-Rosales~\orcidlink{0000-0002-2662-3754}\inst{12} \and
Svetlana Jorstad~\orcidlink{0000-0001-6158-1708}\inst{99} \and
Abhishek V. Joshi~\orcidlink{0000-0002-2514-5965}\inst{37} \and
Taehyun Jung~\orcidlink{0000-0001-7003-8643}\inst{4,48} \and
Mansour Karami~\orcidlink{0000-0001-7387-9333}\inst{44,45} \and
Ramesh Karuppusamy~\orcidlink{0000-0002-5307-2919}\inst{1} \and
Tomohisa Kawashima~\orcidlink{0000-0001-8527-0496}\inst{100} \and
Garrett K. Keating~\orcidlink{0000-0002-3490-146X}\inst{7} \and
Mark Kettenis~\orcidlink{0000-0002-6156-5617}\inst{101} \and
Dong-Jin Kim~\orcidlink{0000-0002-7038-2118}\inst{1} \and
Jae-Young Kim~\orcidlink{0000-0001-8229-7183}\inst{102,1} \and
Jongsoo Kim~\orcidlink{0000-0002-1229-0426}\inst{4} \and
Junhan Kim~\orcidlink{0000-0002-4274-9373}\inst{103} \and
Motoki Kino~\orcidlink{0000-0002-2709-7338}\inst{30,104} \and
Jun Yi Koay~\orcidlink{0000-0002-7029-6658}\inst{34} \and
Prashant Kocherlakota~\orcidlink{0000-0001-7386-7439}\inst{6,7,58} \and
Yutaro Kofuji\inst{86,93} \and
Shoko Koyama~\orcidlink{0000-0002-3723-3372}\inst{105,34} \and
Carsten Kramer~\orcidlink{0000-0002-4908-4925}\inst{43} \and
Joana A. Kramer~\orcidlink{0009-0003-3011-0454}\inst{1} \and
Michael Kramer~\orcidlink{0000-0002-4175-2271}\inst{1} \and
Cheng-Yu Kuo~\orcidlink{0000-0001-6211-5581}\inst{106,34} \and
Noemi La Bella~\orcidlink{0000-0002-8116-9427}\inst{12} \and
Tod R. Lauer~\orcidlink{0000-0003-3234-7247}\inst{107} \and
Daeyoung Lee~\orcidlink{0000-0002-3350-5588}\inst{37} \and
Po Kin Leung~\orcidlink{0000-0002-8802-8256}\inst{108} \and
Aviad Levis~\orcidlink{0000-0001-7307-632X}\inst{42} \and
Zhiyuan Li~\orcidlink{0000-0003-0355-6437}\inst{109,110} \and
Rocco Lico~\orcidlink{0000-0001-7361-2460}\inst{111,2} \and
Greg Lindahl~\orcidlink{0000-0002-6100-4772}\inst{7} \and
Michael Lindqvist~\orcidlink{0000-0002-3669-0715}\inst{20} \and
Jun Liu~\orcidlink{0000-0002-7615-7499}\inst{1} \and
Kuo Liu~\orcidlink{0000-0002-2953-7376}\inst{1} \and
Elisabetta Liuzzo~\orcidlink{0000-0003-0995-5201}\inst{112} \and
Wen-Ping Lo~\orcidlink{0000-0003-1869-2503}\inst{34,113} \and
Laurent Loinard~\orcidlink{0000-0002-5635-3345}\inst{114,6,115} \and
Colin J. Lonsdale~\orcidlink{0000-0003-4062-4654}\inst{5} \and
Amy E. Lowitz~\orcidlink{0000-0002-4747-4276}\inst{8} \and
Nicholas R. MacDonald~\orcidlink{0000-0002-6684-8691}\inst{1} \and
Jirong Mao~\orcidlink{0000-0002-7077-7195}\inst{116,117,118} \and
Nicola Marchili~\orcidlink{0000-0002-5523-7588}\inst{112,1} \and
Sera Markoff~\orcidlink{0000-0001-9564-0876}\inst{119,120} \and
Daniel P. Marrone~\orcidlink{0000-0002-2367-1080}\inst{8} \and
Alan P. Marscher~\orcidlink{0000-0001-7396-3332}\inst{99} \and
Iván Martí-Vidal~\orcidlink{0000-0003-3708-9611}\inst{31,35} \and
Satoki Matsushita~\orcidlink{0000-0002-2127-7880}\inst{34} \and
Lynn D. Matthews~\orcidlink{0000-0002-3728-8082}\inst{5} \and
Lia Medeiros~\orcidlink{0000-0003-2342-6728}\inst{121,11} \and
Karl M. Menten~\orcidlink{0000-0001-6459-0669}\inst{1}\thanks{Deceased} \and
Daniel Michalik~\orcidlink{0000-0002-7618-6556}\inst{122,123} \and
Izumi Mizuno~\orcidlink{0000-0002-7210-6264}\inst{40,41} \and
Yosuke Mizuno~\orcidlink{0000-0002-8131-6730}\inst{60,124,58} \and
James M. Moran~\orcidlink{0000-0002-3882-4414}\inst{6,7} \and
Kotaro Moriyama~\orcidlink{0000-0003-1364-3761}\inst{58,5,86} \and
Monika Moscibrodzka~\orcidlink{0000-0002-4661-6332}\inst{12} \and
Wanga Mulaudzi~\orcidlink{0000-0003-4514-625X}\inst{119} \and
Cornelia Müller~\orcidlink{0000-0002-2739-2994}\inst{1,12} \and
Hendrik Müller~\orcidlink{0000-0002-9250-0197}\inst{125} \and
Alejandro Mus~\orcidlink{0000-0003-0329-6874}\inst{24,25} \and
Gibwa Musoke~\orcidlink{0000-0003-1984-189X}\inst{119,12} \and
Ioannis Myserlis~\orcidlink{0000-0003-3025-9497}\inst{126} \and
Andrew Nadolski~\orcidlink{0000-0001-9479-9957}\inst{29} \and
Hiroshi Nagai~\orcidlink{0000-0003-0292-3645}\inst{30,87} \and
Neil M. Nagar~\orcidlink{0000-0001-6920-662X}\inst{3} \and
Masanori Nakamura~\orcidlink{0000-0001-6081-2420}\inst{127,34} \and
Gopal Narayanan~\orcidlink{0000-0002-4723-6569}\inst{47} \and
Iniyan Natarajan~\orcidlink{0000-0001-8242-4373}\inst{7,6} \and
Antonios Nathanail~\orcidlink{0000-0002-1655-9912}\inst{128,58} \and
Santiago Navarro Fuentes\inst{126} \and
Joey Neilsen~\orcidlink{0000-0002-8247-786X}\inst{129} \and
Roberto Neri~\orcidlink{0000-0002-7176-4046}\inst{43} \and
Chunchong Ni~\orcidlink{0000-0003-1361-5699}\inst{45,46,44} \and
Aristeidis Noutsos~\orcidlink{0000-0002-4151-3860}\inst{1} \and
Michael A. Nowak~\orcidlink{0000-0001-6923-1315}\inst{130} \and
Junghwan Oh~\orcidlink{0000-0002-4991-9638}\inst{101} \and
Hiroki Okino~\orcidlink{0000-0003-3779-2016}\inst{86,93} \and
Héctor R. Olivares Sánchez~\orcidlink{0000-0001-6833-7580}\inst{131} \and
Gisela N. Ortiz-León~\orcidlink{0000-0002-2863-676X}\inst{83,1} \and
Tomoaki Oyama~\orcidlink{0000-0003-4046-2923}\inst{86} \and
Feryal Özel~\orcidlink{0000-0003-4413-1523}\inst{132} \and
Daniel C. M. Palumbo~\orcidlink{0000-0002-7179-3816}\inst{6,7} \and
Jongho Park~\orcidlink{0000-0001-6558-9053}\inst{133,34} \and
Harriet Parsons~\orcidlink{0000-0002-6327-3423}\inst{40,41} \and
Nimesh Patel~\orcidlink{0000-0002-6021-9421}\inst{7} \and
Ue-Li Pen~\orcidlink{0000-0003-2155-9578}\inst{34,44,134,135,136} \and
Dominic W. Pesce~\orcidlink{0000-0002-5278-9221}\inst{7,6} \and
Vincent Piétu~\orcidlink{0009-0006-3497-397X}\inst{43} \and
Richard Plambeck~\orcidlink{0000-0001-6765-9609}\inst{137} \and
Aleksandar PopStefanija~\orcidlink{0000-0002-1889-5450}\inst{47} \and
Oliver Porth~\orcidlink{0000-0002-4584-2557}\inst{119,58} \and
Felix M. Pötzl~\orcidlink{0000-0002-6579-8311}\inst{138} \and
Ben Prather~\orcidlink{0000-0002-0393-7734}\inst{37} \and
Jorge A. Preciado-López~\orcidlink{0000-0002-4146-0113}\inst{44} \and
Giacomo Principe~\orcidlink{0000-0003-0406-7387}\inst{139,140,111} \and
Dimitrios Psaltis~\orcidlink{0000-0003-1035-3240}\inst{132} \and
Hung-Yi Pu~\orcidlink{0000-0001-9270-8812}\inst{141,142,34} \and
Venkatessh Ramakrishnan~\orcidlink{0000-0002-9248-086X}\inst{3,143,18} \and
Ramprasad Rao~\orcidlink{0000-0002-1407-7944}\inst{7} \and
Mark G. Rawlings~\orcidlink{0000-0002-6529-202X}\inst{144,40,41} \and
Angelo Ricarte~\orcidlink{0000-0001-5287-0452}\inst{6,7} \and
Bart Ripperda~\orcidlink{0000-0002-7301-3908}\inst{134,145,135,44} \and
Freek Roelofs~\orcidlink{0000-0001-5461-3687}\inst{12} \and
Alan Rogers~\orcidlink{0000-0003-1941-7458}\inst{5} \and
Cristina Romero-Cañizales~\orcidlink{0000-0001-6301-9073}\inst{34} \and
Arash Roshanineshat~\orcidlink{0000-0002-8280-9238}\inst{8} \and
Helge Rottmann~\orcidlink{0000-0003-1799-8228}\inst{1} \and
Alan L. Roy~\orcidlink{0000-0002-1931-0135}\inst{1} \and
Ignacio Ruiz~\orcidlink{0000-0002-0965-5463}\inst{126} \and
Chet Ruszczyk~\orcidlink{0000-0001-7278-9707}\inst{5} \and
Kazi L. J. Rygl~\orcidlink{0000-0003-4146-9043}\inst{112} \and
Salvador Sánchez~\orcidlink{0000-0002-8042-5951}\inst{126} \and
David Sánchez-Argüelles~\orcidlink{0000-0002-7344-9920}\inst{83,84} \and
Miguel Sánchez-Portal~\orcidlink{0000-0003-0981-9664}\inst{126} \and
Mahito Sasada~\orcidlink{0000-0001-5946-9960}\inst{146,86,147} \and
Kaushik Satapathy~\orcidlink{0000-0003-0433-3585}\inst{8} \and
F. Peter Schloerb\inst{47} \and
Jonathan Schonfeld~\orcidlink{0000-0002-8909-2401}\inst{7} \and
Karl-Friedrich Schuster~\orcidlink{0000-0003-2890-9454}\inst{148} \and
Lijing Shao~\orcidlink{0000-0002-1334-8853}\inst{92,1} \and
Zhiqiang Shen~\orcidlink{0000-0003-3540-8746}\inst{21,54} \and
Des Small~\orcidlink{0000-0003-3723-5404}\inst{101} \and
Bong Won Sohn~\orcidlink{0000-0002-4148-8378}\inst{4,48,55} \and
Jason SooHoo~\orcidlink{0000-0003-1938-0720}\inst{5} \and
León David Sosapanta Salas~\orcidlink{0000-0003-1979-6363}\inst{119} \and
Kamal Souccar~\orcidlink{0000-0001-7915-5272}\inst{47} \and
Joshua S. Stanway~\orcidlink{0009-0003-7659-4642}\inst{149} \and
He Sun~\orcidlink{0000-0003-1526-6787}\inst{150,151} \and
Fumie Tazaki~\orcidlink{0000-0003-0236-0600}\inst{86} \and
Alexandra J. Tetarenko~\orcidlink{0000-0003-3906-4354}\inst{152} \and
Paul Tiede~\orcidlink{0000-0003-3826-5648}\inst{7,6} \and
Michael Titus~\orcidlink{0000-0001-9001-3275}\inst{5} \and
Pablo Torne~\orcidlink{0000-0001-8700-6058}\inst{126,1} \and
Teresa Toscano~\orcidlink{0000-0003-3658-7862}\inst{2} \and
Tyler Trent~\orcidlink{0009-0004-8116-3123}\inst{8} \and
Sascha Trippe~\orcidlink{0000-0003-0465-1559}\inst{153} \and
Matthew Turk~\orcidlink{0000-0002-5294-0198}\inst{29} \and
Ilse van Bemmel~\orcidlink{0000-0001-5473-2950}\inst{68} \and
Huib J. van Langevelde~\orcidlink{0000-0002-0230-5946}\inst{101,15,154} \and
Daniel R. van Rossum~\orcidlink{0000-0001-7772-6131}\inst{12} \and
Jesse Vos~\orcidlink{0000-0003-3349-7394}\inst{12} \and
Jan Wagner~\orcidlink{0000-0003-1105-6109}\inst{1} \and
Derek Ward-Thompson~\orcidlink{0000-0003-1140-2761}\inst{149} \and
John Wardle~\orcidlink{0000-0002-8960-2942}\inst{155} \and
Jasmin E. Washington~\orcidlink{0000-0002-7046-0470}\inst{8} \and
Jonathan Weintroub~\orcidlink{0000-0002-4603-5204}\inst{6,7} \and
Robert Wharton~\orcidlink{0000-0002-7416-5209}\inst{1} \and
Kaj Wiik~\orcidlink{0000-0002-0862-3398}\inst{18,143,156} \and
Gunther Witzel~\orcidlink{0000-0003-2618-797X}\inst{1} \and
Michael F. Wondrak~\orcidlink{0000-0002-6894-1072}\inst{12,157} \and
George N. Wong~\orcidlink{0000-0001-6952-2147}\inst{158,52} \and
Qingwen Wu~\orcidlink{0000-0003-4773-4987}\inst{159} \and
Nitika Yadlapalli~\orcidlink{0000-0003-3255-4617}\inst{42} \and
Paul Yamaguchi~\orcidlink{0000-0002-6017-8199}\inst{7} \and
Aristomenis Yfantis~\orcidlink{0000-0002-3244-7072}\inst{12} \and
Doosoo Yoon~\orcidlink{0000-0001-8694-8166}\inst{119} \and
André Young~\orcidlink{0000-0003-0000-2682}\inst{12} \and
Ken Young~\orcidlink{0000-0002-3666-4920}\inst{7} \and
Ziri Younsi~\orcidlink{0000-0001-9283-1191}\inst{160,58} \and
Wei Yu~\orcidlink{0000-0002-5168-6052}\inst{7} \and
Feng Yuan~\orcidlink{0000-0003-3564-6437}\inst{161} \and
Ye-Fei Yuan~\orcidlink{0000-0002-7330-4756}\inst{162} \and
J. Anton Zensus~\orcidlink{0000-0001-7470-3321}\inst{1} \and
Shuo Zhang~\orcidlink{0000-0002-2967-790X}\inst{163} \and
Guang-Yao Zhao~\orcidlink{0000-0002-4417-1659}\inst{2,1} \and
Shan-Shan Zhao~\orcidlink{0000-0002-9774-3606}\inst{21}}

\institute{
Max-Planck-Institut für Radioastronomie, Auf dem Hügel 69, D-53121 Bonn, Germany \and
Instituto de Astrofísica de Andalucía-CSIC, Glorieta de la Astronomía s/n, E-18008 Granada, Spain \and
Astronomy Department, Universidad de Concepción, Casilla 160-C, Concepción, Chile \and
Korea Astronomy and Space Science Institute, Daedeok-daero 776, Yuseong-gu, Daejeon 34055, Republic of Korea \and
Massachusetts Institute of Technology Haystack Observatory, 99 Millstone Road, Westford, MA 01886, USA \and
Black Hole Initiative at Harvard University, 20 Garden Street, Cambridge, MA 02138, USA \and
Center for Astrophysics $|$ Harvard \& Smithsonian, 60 Garden Street, Cambridge, MA 02138, USA \and
Steward Observatory and Department of Astronomy, University of Arizona, 933 N. Cherry Ave., Tucson, AZ 85721, USA \and
Data Science Institute, University of Arizona, 1230 N. Cherry Ave., Tucson, AZ 85721, USA \and
Program in Applied Mathematics, University of Arizona, 617 N. Santa Rita, Tucson, AZ 85721, USA \and
NASA Hubble Fellowship Program, Einstein Fellow \and
Department of Astrophysics, Institute for Mathematics, Astrophysics and Particle Physics (IMAPP), Radboud University, P.O. Box 9010, 6500 GL Nijmegen, The Netherlands \and
Institute of Astronomy and Astrophysics, Academia Sinica, 645 N. A'ohoku Place, Hilo, HI 96720, USA \and
Department of Physics and Astronomy, University of Hawaii at Manoa, 2505 Correa Road, Honolulu, HI 96822, USA \and
Leiden Observatory, Leiden University, Postbus 2300, 9513 RA Leiden, The Netherlands \and
Netherlands Organisation for Scientific Research (NWO), Postbus 93138, 2509 AC Den Haag, The Netherlands \and
Aalto University Department of Electronics and Nanoengineering, PL 15500, FI-00076 Aalto, Finland \and
Aalto University Metsähovi Radio Observatory, Metsähovintie 114, FI-02540 Kylmälä, Finland \and
National Radio Astronomy Observatory, 520 Edgemont Road, Charlottesville, VA 22903, USA \and
Department of Space, Earth and Environment, Chalmers University of Technology, Onsala Space Observatory, SE-43992 Onsala, Sweden \and
Shanghai Astronomical Observatory, Chinese Academy of Sciences, 80 Nandan Road, Shanghai 200030, People's Republic of China \and
Key Laboratory of Radio Astronomy, Chinese Academy of Sciences, Nanjing 210008, People's Republic of China \and
Instituto de Astronomia, Geofísica e Ciências Atmosféricas, Universidade de São Paulo, R. do Matão, 1226, São Paulo, SP 05508-090, Brazil \and
Dipartimento di Fisica, Università degli Studi di Cagliari, SP Monserrato-Sestu km 0.7, I-09042 Monserrato (CA), Italy \and
INAF - Osservatorio Astronomico di Cagliari, via della Scienza 5, I-09047 Selargius (CA), Italy \and
INFN, sezione di Cagliari, I-09042 Monserrato (CA), Italy \and
Instituto de Física, Pontificia Universidad Católica de Valparaíso, Casilla 4059, Valparaíso, Chile \and
Independent researcher \and
Department of Astronomy, University of Illinois at Urbana-Champaign, 1002 West Green Street, Urbana, IL 61801, USA \and
National Astronomical Observatory of Japan, 2-21-1 Osawa, Mitaka, Tokyo 181-8588, Japan \and
Departament d'Astronomia i Astrofísica, Universitat de València, C. Dr. Moliner 50, E-46100 Burjassot, València, Spain \and
Department of Physics, Faculty of Science, Universiti Malaya, 50603 Kuala Lumpur, Malaysia \and
Department of Physics \& Astronomy, The University of Texas at San Antonio, One UTSA Circle, San Antonio, TX 78249, USA \and
Institute of Astronomy and Astrophysics, Academia Sinica, 11F of Astronomy-Mathematics Building, AS/NTU No. 1, Sec. 4, Roosevelt Rd., Taipei 106216, Taiwan, R.O.C. \and
Observatori Astronòmic, Universitat de València, C. Catedrático José Beltrán 2, E-46980 Paterna, València, Spain \and
Yale Center for Astronomy \& Astrophysics, Yale University, 52 Hillhouse Avenue, New Haven, CT 06511, USA \and
Department of Physics, University of Illinois, 1110 West Green Street, Urbana, IL 61801, USA \and
Fermi National Accelerator Laboratory, MS209, P.O. Box 500, Batavia, IL 60510, USA \and
Department of Astronomy and Astrophysics, University of Chicago, 5640 South Ellis Avenue, Chicago, IL 60637, USA \and
East Asian Observatory, 660 N. A'ohoku Place, Hilo, HI 96720, USA \and
James Clerk Maxwell Telescope (JCMT), 660 N. A'ohoku Place, Hilo, HI 96720, USA \and
California Institute of Technology, 1200 East California Boulevard, Pasadena, CA 91125, USA \and
Institut de Radioastronomie Millimétrique (IRAM), 300 rue de la Piscine, F-38406 Saint Martin d'Hères, France \and
Perimeter Institute for Theoretical Physics, 31 Caroline Street North, Waterloo, ON N2L 2Y5, Canada \and
Department of Physics and Astronomy, University of Waterloo, 200 University Avenue West, Waterloo, ON N2L 3G1, Canada \and
Waterloo Centre for Astrophysics, University of Waterloo, Waterloo, ON N2L 3G1, Canada \and
Department of Astronomy, University of Massachusetts, Amherst, MA 01003, USA \and
University of Science and Technology, Gajeong-ro 217, Yuseong-gu, Daejeon 34113, Republic of Korea \and
Kavli Institute for Cosmological Physics, University of Chicago, 5640 South Ellis Avenue, Chicago, IL 60637, USA \and
Department of Physics, University of Chicago, 5720 South Ellis Avenue, Chicago, IL 60637, USA \and
Enrico Fermi Institute, University of Chicago, 5640 South Ellis Avenue, Chicago, IL 60637, USA \and
Princeton Gravity Initiative, Jadwin Hall, Princeton University, Princeton, NJ 08544, USA \and
Cornell Center for Astrophysics and Planetary Science, Cornell University, Ithaca, NY 14853, USA \and
Key Laboratory of Radio Astronomy and Technology, Chinese Academy of Sciences, A20 Datun Road, Chaoyang District, Beijing, 100101, People's Republic of China \and
Department of Astronomy, Yonsei University, Yonsei-ro 50, Seodaemun-gu, 03722 Seoul, Republic of Korea \and
Physics Department, Fairfield University, 1073 North Benson Road, Fairfield, CT 06824, USA \and
Instituto de Astronomía, Universidad Nacional Autónoma de México (UNAM), Apdo Postal 70-264, Ciudad de México, México \and
Institut für Theoretische Physik, Goethe-Universität Frankfurt, Max-von-Laue-Straße 1, D-60438 Frankfurt am Main, Germany \and
Research Center for Astronomical Computing, Zhejiang Laboratory, Hangzhou 311100, People's Republic of China \and
Tsung-Dao Lee Institute, Shanghai Jiao Tong University, Shengrong Road 520, Shanghai, 201210, People's Republic of China \and
Department of Astronomy and Columbia Astrophysics Laboratory, Columbia University, 500 W. 120th Street, New York, NY 10027, USA \and
Center for Computational Astrophysics, Flatiron Institute, 162 Fifth Avenue, New York, NY 10010, USA \and
Dipartimento di Fisica ``E. Pancini'', Università di Napoli ``Federico II'', Compl. Univ. di Monte S. Angelo, Edificio G, Via Cinthia, I-80126, Napoli, Italy \and
INFN Sez. di Napoli, Compl. Univ. di Monte S. Angelo, Edificio G, Via Cinthia, I-80126, Napoli, Italy \and
Wits Centre for Astrophysics, University of the Witwatersrand, 1 Jan Smuts Avenue, Braamfontein, Johannesburg 2050, South Africa \and
Department of Physics, University of Pretoria, Hatfield, Pretoria 0028, South Africa \and
Centre for Radio Astronomy Techniques and Technologies, Department of Physics and Electronics, Rhodes University, Makhanda 6140, South Africa \and
ASTRON, Oude Hoogeveensedijk 4, 7991 PD Dwingeloo, The Netherlands \and
LESIA, Observatoire de Paris, Université PSL, CNRS, Sorbonne Université, Université de Paris, 5 place Jules Janssen, F-92195 Meudon, France \and
JILA and Department of Astrophysical and Planetary Sciences, University of Colorado, Boulder, CO 80309, USA \and
National Astronomical Observatories, Chinese Academy of Sciences, 20A Datun Road, Chaoyang District, Beijing 100101, PR China \and
Las Cumbres Observatory, 6740 Cortona Drive, Suite 102, Goleta, CA 93117-5575, USA \and
Department of Physics, University of California, Santa Barbara, CA 93106-9530, USA \and
Department of Electrical Engineering and Computer Science, Massachusetts Institute of Technology, 32-D476, 77 Massachusetts Ave., Cambridge, MA 02142, USA \and
Google Research, 355 Main St., Cambridge, MA 02142, USA \and
Institut für Theoretische Physik und Astrophysik, Universität Würzburg, Emil-Fischer-Str. 31, \and
Department of History of Science, Harvard University, Cambridge, MA 02138, USA \and
Department of Physics, Harvard University, Cambridge, MA 02138, USA \and
NCSA, University of Illinois, 1205 W. Clark St., Urbana, IL 61801, USA \and
Institute for Mathematics and Interdisciplinary Center for Scientific Computing, Heidelberg University, Im Neuenheimer Feld 205, Heidelberg 69120, Germany \and
Institut f\"ur Theoretische Physik, Universit\"at Heidelberg, Philosophenweg 16, 69120 Heidelberg, Germany \and
CP3-Origins, University of Southern Denmark, Campusvej 55, 5230 Odense, Denmark \and
Instituto Nacional de Astrofísica, Óptica y Electrónica. Apartado Postal 51 y 216, 72000. Puebla Pue., México \and
Consejo Nacional de Humanidades, Ciencia y Tecnología, Av. Insurgentes Sur 1582, 03940, Ciudad de México, México \and
Key Laboratory for Research in Galaxies and Cosmology, Chinese Academy of Sciences, Shanghai 200030, People's Republic of China \and
Mizusawa VLBI Observatory, National Astronomical Observatory of Japan, 2-12 Hoshigaoka, Mizusawa, Oshu, Iwate 023-0861, Japan \and
Department of Astronomical Science, The Graduate University for Advanced Studies (SOKENDAI), 2-21-1 Osawa, Mitaka, Tokyo 181-8588, Japan \and
Department of Physics, McGill University, 3600 rue University, Montréal, QC H3A 2T8, Canada \and
Trottier Space Institute at McGill, 3550 rue University, Montréal,  QC H3A 2A7, Canada \and
NOVA Sub-mm Instrumentation Group, Kapteyn Astronomical Institute, University of Groningen, Landleven 12, 9747 AD Groningen, The Netherlands \and
Department of Astronomy, School of Physics, Peking University, Beijing 100871, People's Republic of China \and
Kavli Institute for Astronomy and Astrophysics, Peking University, Beijing 100871, People's Republic of China \and
Department of Astronomy, Graduate School of Science, The University of Tokyo, 7-3-1 Hongo, Bunkyo-ku, Tokyo 113-0033, Japan \and
The Institute of Statistical Mathematics, 10-3 Midori-cho, Tachikawa, Tokyo, 190-8562, Japan \and
Department of Statistical Science, The Graduate University for Advanced Studies (SOKENDAI), 10-3 Midori-cho, Tachikawa, Tokyo 190-8562, Japan \and
Kavli Institute for the Physics and Mathematics of the Universe, The University of Tokyo, 5-1-5 Kashiwanoha, Kashiwa, 277-8583, Japan \and
ASTRAVEO LLC, PO Box 1668, Gloucester, MA 01931 \and
Applied Materials Inc., 35 Dory Road, Gloucester, MA 01930  \and
Institute for Astrophysical Research, Boston University, 725 Commonwealth Ave., Boston, MA 02215, USA \and
Institute for Cosmic Ray Research, The University of Tokyo, 5-1-5 Kashiwanoha, Kashiwa, Chiba 277-8582, Japan \and
Joint Institute for VLBI ERIC (JIVE), Oude Hoogeveensedijk 4, 7991 PD Dwingeloo, The Netherlands \and
Department of Physics, Ulsan National Institute of Science and Technology (UNIST), 50 UNIST-gil, Eonyang-eup, Ulju-gun, Ulsan 44919, Republic of Korea \and
Department of Physics, Korea Advanced Institute of Science and Technology (KAIST), 291 Daehak-ro, Yuseong-gu, Daejeon 34141, Republic of Korea \and
Kogakuin University of Technology \& Engineering, Academic Support Center, 2665-1 Nakano, Hachioji, Tokyo 192-0015, Japan \and
Graduate School of Science and Technology, Niigata University, 8050 Ikarashi 2-no-cho, Nishi-ku, Niigata 950-2181, Japan \and
Physics Department, National Sun Yat-Sen University, No. 70, Lien-Hai Road, Kaosiung City 80424, Taiwan, R.O.C. \and
National Optical Astronomy Observatory, 950 N. Cherry Ave., Tucson, AZ 85719, USA \and
Department of Physics, The Chinese University of Hong Kong, Shatin, N. T., Hong Kong \and
School of Astronomy and Space Science, Nanjing University, Nanjing 210023, People's Republic of China \and
Key Laboratory of Modern Astronomy and Astrophysics, Nanjing University, Nanjing 210023, People's Republic of China \and
INAF-Istituto di Radioastronomia, Via P. Gobetti 101, I-40129 Bologna, Italy \and
INAF-Istituto di Radioastronomia \& Italian ALMA Regional Centre, Via P. Gobetti 101, I-40129 Bologna, Italy \and
Department of Physics, National Taiwan University, No. 1, Sec. 4, Roosevelt Rd., Taipei 106216, Taiwan, R.O.C \and
Instituto de Radioastronomía y Astrofísica, Universidad Nacional Autónoma de México, Morelia 58089, México \and
David Rockefeller Center for Latin American Studies, Harvard University, 1730 Cambridge Street, Cambridge, MA 02138, USA \and
Yunnan Observatories, Chinese Academy of Sciences, 650011 Kunming, Yunnan Province, People's Republic of China \and
Center for Astronomical Mega-Science, Chinese Academy of Sciences, 20A Datun Road, Chaoyang District, Beijing, 100012, People's Republic of China \and
Key Laboratory for the Structure and Evolution of Celestial Objects, Chinese Academy of Sciences, 650011 Kunming, People's Republic of China \and
Anton Pannekoek Institute for Astronomy, University of Amsterdam, Science Park 904, 1098 XH, Amsterdam, The Netherlands \and
Gravitation and Astroparticle Physics Amsterdam (GRAPPA) Institute, University of Amsterdam, Science Park 904, 1098 XH Amsterdam, The Netherlands \and
Department of Astrophysical Sciences, Peyton Hall, Princeton University, Princeton, NJ 08544, USA \and
Science Support Office, Directorate of Science, European Space Research and Technology Centre (ESA/ESTEC), Keplerlaan 1, 2201 AZ Noordwijk, The Netherlands \and
Department of Astronomy and Astrophysics, University of Chicago, \and
School of Physics and Astronomy, Shanghai Jiao Tong University, 800 Dongchuan Road, Shanghai, 200240, People's Republic of China \and
National Radio Astronomy Observatory, P.O. Box O, Socorro, NM 87801, USA \and
Institut de Radioastronomie Millimétrique (IRAM), Avenida Divina Pastora 7, Local 20, E-18012, Granada, Spain \and
National Institute of Technology, Hachinohe College, 16-1 Uwanotai, Tamonoki, Hachinohe City, Aomori 039-1192, Japan \and
Research Center for Astronomy, Academy of Athens, Soranou Efessiou 4, 115 27 Athens, Greece \and
Department of Physics, Villanova University, 800 Lancaster Avenue, Villanova, PA 19085, USA \and
Physics Department, Washington University, CB 1105, St. Louis, MO 63130, USA \and
Departamento de Matemática da Universidade de Aveiro and Centre for Research and Development in Mathematics and Applications (CIDMA), Campus de Santiago, 3810-193 Aveiro, Portugal \and
School of Physics, Georgia Institute of Technology, 837 State St NW, Atlanta, GA 30332, USA \and
School of Space Research, Kyung Hee University, 1732, Deogyeong-daero, Giheung-gu, Yongin-si, Gyeonggi-do 17104, Republic of Korea \and
Canadian Institute for Theoretical Astrophysics, University of Toronto, 60 St. George Street, Toronto, ON M5S 3H8, Canada \and
Dunlap Institute for Astronomy and Astrophysics, University of Toronto, 50 St. George Street, Toronto, ON M5S 3H4, Canada \and
Canadian Institute for Advanced Research, 180 Dundas St West, Toronto, ON M5G 1Z8, Canada \and
Radio Astronomy Laboratory, University of California, Berkeley, CA 94720, USA \and
Institute of Astrophysics, Foundation for Research and Technology - Hellas, Voutes, 7110 Heraklion, Greece \and
Dipartimento di Fisica, Università di Trieste, I-34127 Trieste, Italy \and
INFN Sez. di Trieste, I-34127 Trieste, Italy \and
Department of Physics, National Taiwan Normal University, No. 88, Sec. 4, Tingzhou Rd., Taipei 116, Taiwan, R.O.C. \and
Center of Astronomy and Gravitation, National Taiwan Normal University, No. 88, Sec. 4, Tingzhou Road, Taipei 116, Taiwan, R.O.C. \and
Finnish Centre for Astronomy with ESO, FI-20014 University of Turku, Finland \and
Gemini Observatory/NSF NOIRLab, 670 N. A’ohōkū Place, Hilo, HI 96720, USA \and
Department of Physics, University of Toronto, 60 St. George Street, Toronto, ON M5S 1A7, Canada \and
Department of Physics, Tokyo Institute of Technology, 2-12-1 Ookayama, Meguro-ku, Tokyo 152-8551, Japan \and
Hiroshima Astrophysical Science Center, Hiroshima University, 1-3-1 Kagamiyama, Higashi-Hiroshima, Hiroshima 739-8526, Japan \and
Institut de Radioastronomie Millimétrique (IRAM), 300 rue de la Piscine, \and
Jeremiah Horrocks Institute, University of Central Lancashire, Preston PR1 2HE, UK \and
National Biomedical Imaging Center, Peking University, Beijing 100871, People's Republic of China \and
College of Future Technology, Peking University, Beijing 100871, People's Republic of China \and
Department of Physics and Astronomy, University of Lethbridge, Lethbridge, Alberta T1K 3M4, Canada \and
Department of Physics and Astronomy, Seoul National University, Gwanak-gu, Seoul 08826, Republic of Korea \and
University of New Mexico, Department of Physics and Astronomy, Albuquerque, NM 87131, USA \and
Physics Department, Brandeis University, 415 South Street, Waltham, MA 02453, USA \and
Tuorla Observatory, Department of Physics and Astronomy, FI-20014 University of Turku, Finland \and
Radboud Excellence Fellow of Radboud University, Nijmegen, The Netherlands \and
School of Natural Sciences, Institute for Advanced Study, 1 Einstein Drive, Princeton, NJ 08540, USA \and
School of Physics, Huazhong University of Science and Technology, Wuhan, Hubei, 430074, People's Republic of China \and
Mullard Space Science Laboratory, University College London, Holmbury St. Mary, Dorking, Surrey, RH5 6NT, UK \and
Center for Astronomy and Astrophysics and Department of Physics, Fudan University, Shanghai 200438, People's Republic of China \and
Astronomy Department, University of Science and Technology of China, Hefei 230026, People's Republic of China \and
Department of Physics and Astronomy, Michigan State University, 567 Wilson Rd, East Lansing, MI 48824, USA}

\date{Draft; \today}

\abstract
{
 The 2017 observing campaign of the Event Horizon Telescope (EHT) delivered the first very long baseline interferometry (VLBI) images at the observing frequency of 230\,GHz, leading to a number of unique studies on black holes and relativistic jets from active galactic nuclei (AGN). In total, eighteen sources were observed, including the main science targets, Sgr\,A* and M\,87 and various calibrators. Sixteen sources were AGN.
 }
{ 
 We investigated the morphology of the sixteen AGN in the EHT 2017 data set, focusing on the properties of the VLBI cores: size, flux density, and brightness temperature. We studied their dependence on the observing frequency in order to compare it with the Blandford-K\"onigl~(BK) jet model. In particular, we aimed to study the signatures of jet acceleration and magnetic energy conversion. 
}
{
We modeled the source structure of seven AGN in the EHT 2017 data set using linearly polarized circular Gaussian components (1749+096, 1055+018, BL\,Lac, J0132--1654, J0006--0623, CTA\,102, 3C\,454.3) and collected results for the other nine AGN from dedicated EHT publications, complemented by lower frequency data in the 2-86\,GHz range. Combining these data into a multi-frequency EHT+ data set, we studied the dependences of the VLBI core component flux density, size, and brightness temperature on the frequency measured in the AGN host frame (and hence on the distance from the central black hole), characterizing them with power law fits. We compared the observations with the BK jet model and estimated the magnetic field strength dependence on the distance from the central black hole.
}
{
 Our observations spanning event horizon- to parsec scales indicate a deviation from the standard BK model, particularly in the 
 decrease of the brightness temperature with the observing frequency. Only some of the discrepancies may be alleviated by tweaking the model parameters or the jet collimation 
 profile. Either bulk acceleration of the jet material, energy transfer from the magnetic field to the particles, or both are required to explain the observations. For our sample, we estimate a general radial dependence of the Doppler factor $\delta \propto r^{\leq0.5}$. This interpretation is consistent with a~magnetically accelerated sub-parsec jet. We also estimate a steep decrease of the magnetic field strength with radius $B \propto r^{-3}$, hinting at jet acceleration or efficient magnetic energy dissipation.
}
{}
\maketitle
%
\section{Introduction}

Some supermassive black holes (SMBHs) in the centers of galaxies power highly collimated outflows reaching thousands of parsecs into interstellar, and sometimes intergalactic space: relativistic jets. 
These jets are formed and accelerated in a compact region close to the black hole, the active galactic nucleus (AGN).
The detailed physics governing this region, such as the mechanisms for jet formation, collimation and acceleration, as well as 
the role of magnetic fields, are subjects of extensive active work \citep[see, e.\,g.,][for a review of the current state of AGN jet research]{Blandford2019}.

The central nuclear region can be resolved with very long baseline interferometry (VLBI) radio observations \citep[see][for a review]{Boccardi2017}. 
In images obtained from such observations, AGN often 
show a "core-jet" structure, with a bright feature referred to as the VLBI core, and
a lower-intensity extended jet. The VLBI core corresponds to the synchrotron photosphere of the outgoing jet flow, that is, a transition between an optically thick inner region and an optically thin outer region. Thus, its location and properties depend on the 
observing frequency, with cores observed at higher frequencies approaching the central engine of the system \citep["core shift", e.\,g.,][]{Pushkarev2012MOJAVE}. This way, multi-frequency radio-interferometric observations enable studies of the system's properties along the jet, constraining the energy contents of
particles and magnetic fields, and indicating the acceleration mechanism.

\begin{table*}[htp!]
\setlength{\tabcolsep}{5pt}
\caption{Summary of the EHT 2017 observations at 230\,GHz}
\begin{center}
\begin{tabular}{llllllllll}
\hline
\hline
 Common & B1950 & J2000 & Redshift $z$ & Type & Volume & Days (Apr 2017) & Sites  & BL (km) & Ref.\\ \hline
Sagittarius\,A$^*$ & 1742--286  & J1750--2900  & -- & -- & 23024 & 5, 6, 7, 10, 11  & all & 11204 & (1)\\ 
3C\,279 & 1253--055 & J1256--0547 & 0.538 & Q & 9638 & 5, 6, 10, 11 &  all  & 11297 & (2) \\ 
M\,87 & 1228+126 & J1230+1223 & 0.0042 & R & 8846  & 5, 6, 10, 11 & all but Y & 10835 & (3) \\ 
OJ\,287 & 0851+202 & J0854+2006 & 0.306 & B & 4548 & 5, 10, 11 & all but Y & 10834 & (4)  \\ 
Centaurus\,A & 1322--427 & J1325-4303  & 0.0014 & R   & 3218 & 10 & all but P & 10749 & (5) \\ 
J1924--2914 & 1921--293 & J1924--2914 & 0.353 & Q & 2937  & 5, 6, 7, 10, 11 &  all & 11091 & (6) \\ 
NRAO\,530 & 1730--130 & J1733--1304 & 0.902 & Q   & 1439 & 5, 6, 7 & all & 11305 & (7) \\ 
3C\,273 & 1226+023 & J1229+0203 & 0.158 & Q & 642  & 6 & all but Y & 9372 & (8)   \\ 
1749+096 & 1749+096 & J1751+0939 & 0.322 & B  & 612 & 10 & J, L, S, X, Z &  9342 & \\ 
1055+018 & 1055+018 & J1058+0133 & 0.89 & Q & 488 & 5, 10, 11 & \jref{all but Y} & 8904 &  \\ 
BL\,Lacertae & 2200+420 & J2202+4216 & 0.069 & B & 174 & 7 & J, S, X, Z & 8934 &  \\ 
3C\,84 & 0316+413 & J0319+4130 & 0.0177 & R & 141  & 7 & A, J, S, X, Z & 9344 & (9)\\ 
3C\,454.3 & 2251+158 & J2253+1608 & 0.859 & Q &  139 &  7 &  J, S, X & 9382 &\\  
CTA\,102 & 2230+114 & J2232+1143 & 1.037 & Q & 118  & 7 & J, S, X & 9384 & \\ 
J0132--1654 & 0130--171 & J0132--1654 & 1.020 & Q  & 65 & 6, 7 & A, J, P, Y & 11263 & \\ 
J0006--0623 & 0003--066 &  J0006--0623 & 0.347 & B  & 36 & 7 & A, J, P, Y & 11211 & \\ 
NGC\,1052 & 0238--084 & J0241--0815 & 0.0049 & R  & 20 & 6, 7 & A, J, S, Y & 5895  & (10)\\ 
Cygnus\,X--3 &  2030+407 & J2032+4057   & -- &  X  & 16 & 5 & J, S & $< 0.2$ &   \\
\hline  
\end{tabular}
\end{center}
\vspace{-0.3cm}
\tablefoot{The names listed in column "Common" follow the convention used by the EHT, and sometimes coincide with the "B1950" or "J2000" designations. 
Object types: B: BL\,Lac object; Q: quasar; R: radio galaxy; X: X-ray binary. 
Sites: A = ALMA;  J = JCMT; L = LMT; P = PV; S = SMA; X = APEX; Y = SPT; Z = SMT.  
Redshifts taken from SIMBAD \citep{Wenger2000}; Sgr\,A* and Cyg X--3 are galactic sources.
Volume: Total number of detections; BL: Maximum projected baseline.
}
\tablebib{
(1) \cite{SgraP1}; 
(2) \cite{Kim2020}; 
(3) \cite{M87p1}; 
(4) Gomez et al. in prep.;
(5) \cite{Janssen2021};
(6) \cite{Issaoun2022};
(7) \cite{Jorstad2023}; 
(8) Wielgus et al. in prep.;
(9) \cite{Paraschos2024};
(10) \cite{Baczko2024}
}
\label{tab:source_list}
\end{table*}

The observing campaign of the Event~Horizon~Telescope~(EHT) in April 2017 led to the first VLBI images of supermassive black holes.
The primary event horizon scale science targets of the EHT were M\,87* \citep{M87p1,M87p2,M87p3,M87p4,M87p5,M87p6,M87p7,M87p8,M87P9}, and 
Sagittarius~A* \citep{SgraP1,SgraP2,SgraP3,SgraP4,SgraP5,SgraP6,SgraP7,SgraP8}. Additionally, new results on several AGN were obtained: 3C\,279 \citep{Kim2020}, Centaurus\,A \citep{Janssen2021}, J1924--2914 
\citep{Issaoun2022}, NRAO\,530 \citep{Jorstad2023}, 3C\,84 \citep{Paraschos2024}, and NGC\,1052 \citep{Baczko2024}. While the efforts to analyze individual EHT targets continue (J.~L. Gomez et al. in prep., M. Wielgus et al. in prep.), the first conclusions about the statistical properties of AGN sources at 230\,GHz can be drawn, based on the EHT 2017 sample.

\begin{figure*}[h!]
	\centering
	\includegraphics[width=0.995\textwidth]{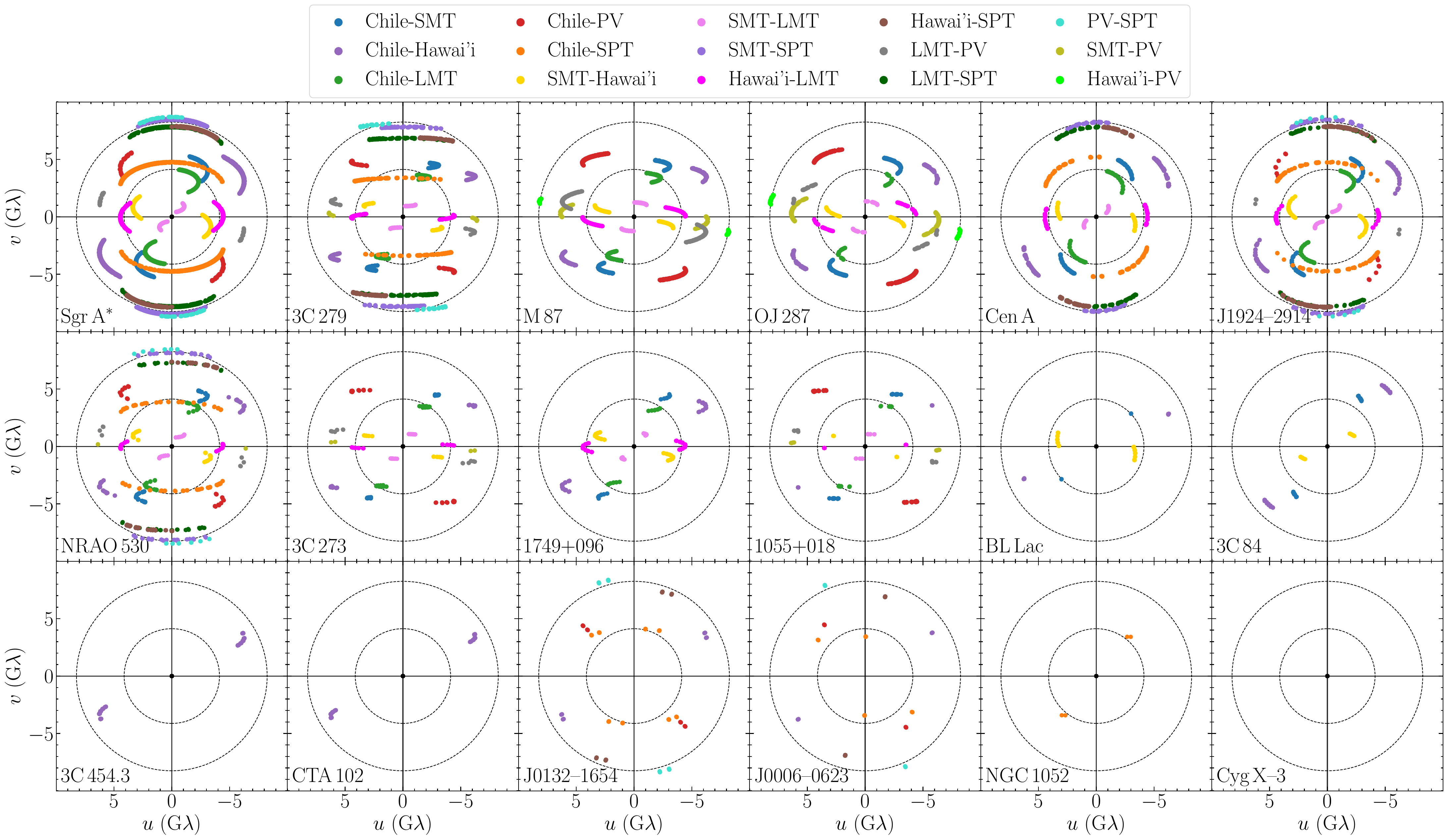} 
    \caption{$(u,v)$-coverage for all sources observed during the EHT 2017 campaign, as summarized in Table \ref{tab:source_list}, combining all available visibility detections for all days and bands, averaged in 120\,s intervals. The two circles in each panel correspond to the fringe spacing characterizing an instrumental resolution of 25 and 50\,$\upmu$as. JCMT-SMA and ALMA-APEX are the very short intrasite baselines, shown as black data points near  the origin of the $(u,v)$ coordinate system (when available). JCMT and SMA are shown as "Hawai'i", while ALMA and APEX are shown as "Chile".}
	\label{fig:UVcoverage}
\end{figure*}

During the 2017 campaign, the EHT observed a total number of eighteen different sources, achieving detections on exceptionally long baselines (over 4.5\,G$\lambda$) for seventeen objects; see Table~\ref{tab:source_list} 
for a summary of the observations. The set includes EHT collaboration projects (M\,87, Sgr~A$^*$), projects proposed by individual researchers (3C\,279, OJ\,287, Cen\,A and others), along with all calibrator sources used (J1924--2914, NRAO\,530, 3C\,273 and others).
While the $(u,v)$-coverage for many of these sources is insufficient for imaging, as illustrated in Fig.~\ref{fig:UVcoverage}, it is possible to estimate their angular size, brightness temperature, and fractional polarization. Such physical parameters are of particular importance to constrain theoretical models of accretion onto and outflows from the vicinity of black holes  
\citep[e.\,g.,][]{Blandford1979,Blandford1982,Gabuzda2017,Gabuzda2018,MacDonald2018,Kramer2021,Cruz-Osorio2022,Fromm2022,Roder2023}.

With this work, we inaugurate a 230\,GHz catalogue of sources that will grow with subsequent EHT campaigns. In particular, this work adds 230\,GHz observations to the large existing sample of sources from surveys at lower frequencies. Since the mid-1990s, 86\,GHz surveys have been carried out, first with the Coordinate 
Millimeter VLBI Array \citep[CMVA;][]{Rogers1995,Beasley1997,Rantakyro1998,Lonsdale1998,Lobanov2000,Lee2008}, which was 
ultimately succeeded by the Global Millimeter VLBI Array \citep[GMVA;][]{Lee2008,Nair2019}. \cite{Pushkarev2012} carried out 
a~survey at 2\,GHz and 8\,GHz using a~combination of the Very Long Baseline Array (VLBA) and up to ten geodetic antennas. The 
long-running MOJAVE\footnote{Monitoring of jets in active galactic nuclei with VLBA experiments} \citep{Kellermann2004,Lister2005,Kovalev2005,Homan2006,Cohen2007,Lister2009,Lister2018,Lister2021} and VLBA-BU-BLAZAR/BEAM-ME\footnote{Blazars entering the astrophysical multi-messenger era} 
programs \citep{Jorstad2017,Weaver2022} supply data at intermediate frequencies 15\,GHz and 43\,GHz, respectively, observed 
with the VLBA. Surveys at 5\,GHz have been carried out with the 
VLBA in the frame of the VLBA imaging and polarimetry survey \citep[VIPS;][]{helmboldt2007,Helmboldt2008} and the VLBI space observatory programme \citep[VSOP; e.\,g., ][]{Dodson2008}.


Characteristic properties of AGN jets may be revealed using a statistical approach to the investigation of the brightness 
temperature and its dependence on frequency, and in turn, on the distance from the central engine \citep{Blandford1979}. 
Given a sufficiently large sample size, 
such an approach remains robust against uncertainties related to properties of individual sources, such as poorly 
constrained inclination and bulk flow velocity, as long as errors are uncorrelated.
Adding measurements at higher frequencies is expected to 
greatly enhance the results of such statistical analyses, extending the investigated jet region closer to the supermassive 
black hole (SMBH) and thus allowing for more accurate tests of the inner jet models, including their launching \citep[e.g.,][]{blandfordznajeck1977, Blandford1982} and acceleration in the vicinity of the true central AGN engine \citep[e.g.,][]{Blandford1979, Marscher1995,heinz2000,vlahakis2004}. 

Throughout this paper we adopt a cosmology
with $H_0 = 67.7 \, \textrm{km} \, \textrm{s}^{-1} \, \textrm{Mpc}^{-1}$, $\Omega_m = 0.307$, and $\Omega_\Lambda = 0.693$ \citep{Planck}.

\section{EHT results}

\subsection{EHT observations and data reduction}

Eight facilities participated in the EHT observing campaign on April 5-11, 2017: The Atacama Large Millimeter/submillimeter Array \citep[ALMA, A, operating 
as a phased array;][]{Matthews2018, Goddi2019} and the Atacama Pathfinder 
Experiment (APEX, X) telescopes in Chile; the Large Millimeter Telescope 
Alfonso Serrano (LMT, L) in Mexico; the IRAM 30\,m telescope (PV, P) in Spain; 
the Submillimeter Telescope (SMT, Z) in Arizona; the James Clerk Maxwell 
Telescope (JCMT, J) and the Submillimeter Array (SMA, S) in Hawai'i; and the 
South Pole Telescope (SPT, Y) in Antarctica. Two frequency bands, each 2\,GHz 
wide, centered at 227.1\,GHz (LO band) and 229.1\,GHz (HI band) were 
recorded. For a detailed description of the EHT array instrumental 
configuration see \citet{M87p2}. 

Following correlation, the data were 
reduced using the \texttt{EHT-HOPS} \citep[][]{Blackburn2019} and 
\texttt{rPICARD} \citep{Janssen2019} pipelines to independently 
validate the results \citep{M87p3}. The EHT calibration procedures are 
described in detail in \citet{SgraP2}, with minor updates with respect 
to \citet{M87p3}. Whenever applicable, polarization leakage was calibrated following 
\citet{M87p7} and \citet{Issaoun2022}. \jref{The electric vector position angle (EVPA) calibration requires persistent participation of ALMA in the EHT observing array. As a consequence, the polarization leakage calibration could only be applied in a straightforward way to sources with coverage as good or better than that of 3C\,273 (see Table~\ref{tab:source_list} and Fig.~\ref{fig:UVcoverage}, as well as \citealt{Paraschos2024} for further details). In particular, the seven sources introduced in this paper (Table~\ref{tab:source_list}) were not calibrated for polarization leakage and the absolute EVPA. This issue has minimal impact on the total intensity analysis, but the polarimetric analysis is affected, see Appendix~\ref{app:modelfits}.} 
While EHT resolves structures on $\sim$10--500\,$\upmu$as angular scales, simultaneous ALMA-only measurements of flux densities and fractional polarization at 212--230\,GHz 
at angular resolutions of $\sim$\,1\,arcsec have been reported for a number of 
observed sources by \citet{Goddi2021}, constraining the total flux density of the core and the compact jet.

\subsection{EHT data sets and model fitting}
\label{subsec:modeling}

A summary of the EHT observations in April 2017 is given in Table~\ref{tab:source_list}. The "Volume" column contains the number of scan-averaged detected visibilities (that is, time-averaged for several minutes, depending on a particular schedule, with separate polarimetric correlation products counted individually), indicating the 
relative constraining power of the respective data sets; see also Fig.~\ref{fig:UVcoverage} for a comparison of the $(u,v)-$coverage between the different EHT data sets. For Cygnus~X--3 we only measured a~short (intrasite) baseline flux density, preventing us from estimating source 
compactness. Apart from the two galactic sources (Sgr~A$^*$ and Cyg~X--3), the EHT data set contains observations of sixteen AGN sources, which are the subject of the analysis presented in this paper. For these sources we provide estimates of the black hole masses and Doppler factors in Appendix~\ref{Appendix:Doppler}.

Whenever a dedicated study of an individual EHT target is available ("Ref." column in Table~\ref{tab:source_list}), we used source parameters reported therein. For the 
remaining seven AGN sources with very sparse $(u,v)$--coverage and with no 
dedicated analysis published, we performed geometric model 
fitting with linearly polarized circular Gaussian components using 
\texttt{eht-imaging} \citep{Chael2016,ehtim2022,Roelofs2023}, exploiting 
heuristic optimisation tools implemented in \texttt{SciPy} 
\citep{2020SciPy} to search for the best-fit solution. For each of these 
sources we used all available data (LO and HI bands, all available days) to 
constrain a single geometric model. The number of polarized circular 
Gaussian components was chosen based on the minimal number of the model 
degrees of freedom required to obtain a high quality fit to visibility 
amplitudes, closure phases, and fractional linear polarizations, generally 
characterized by the reduced $\chi$-square $\chi_n^2<2$. 

\begin{figure*}[h!]
	\centering
    \hspace{0.02cm}\includegraphics[width=5.25cm]{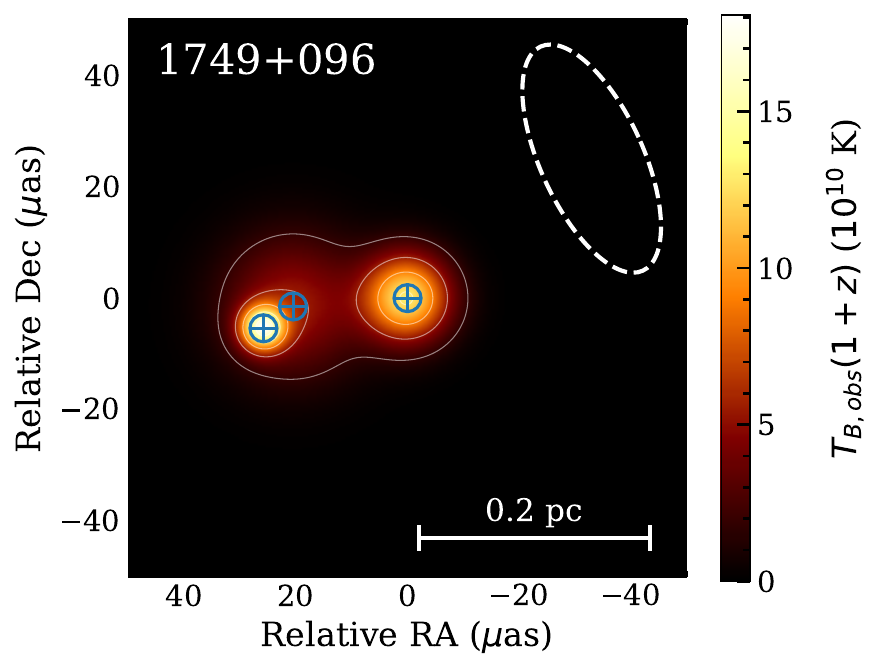}\hspace{0.02cm}
    \includegraphics[width=10.5cm]{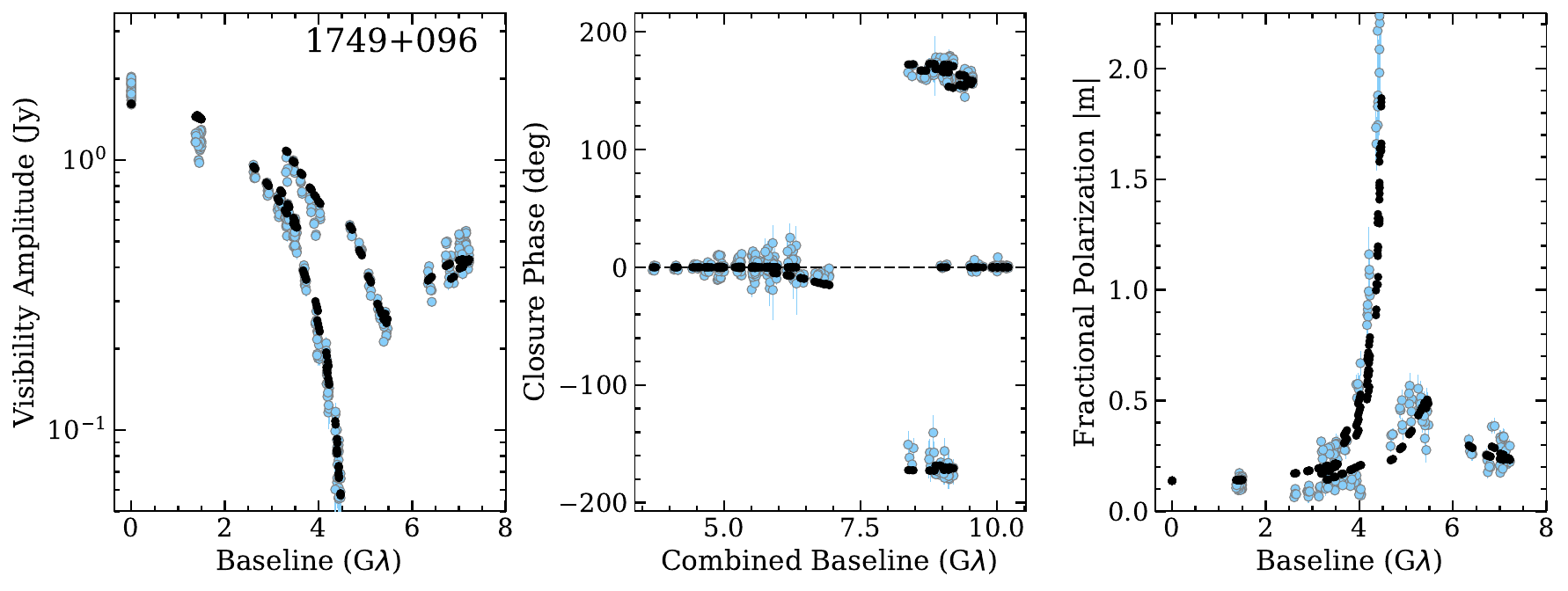}

    \includegraphics[width=5.35cm]{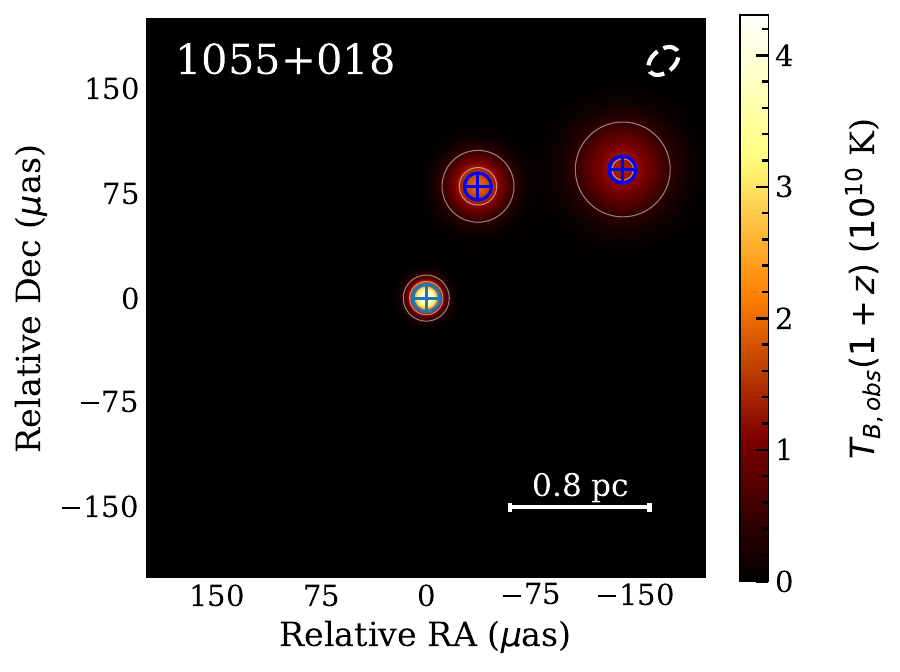}\hspace{-0.025cm}
    \includegraphics[width=10.6cm]{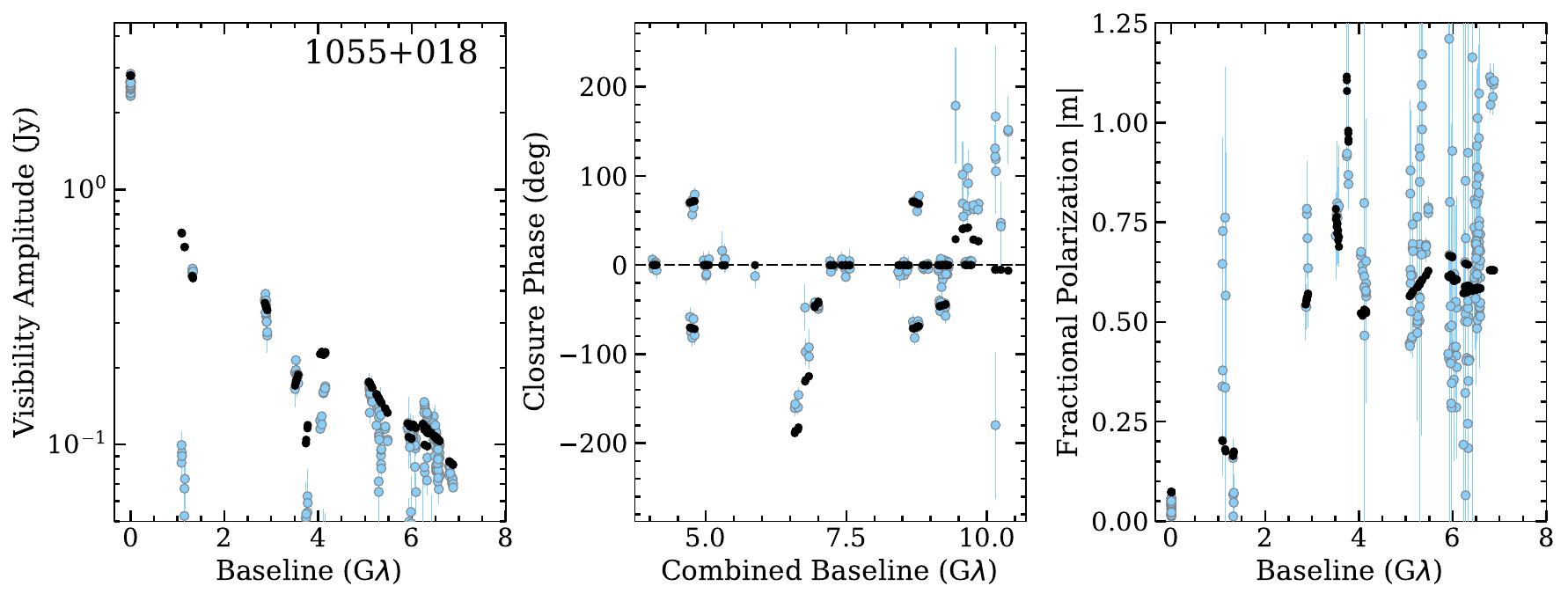}

    \hspace{0.09cm}\includegraphics[width=5.25cm]{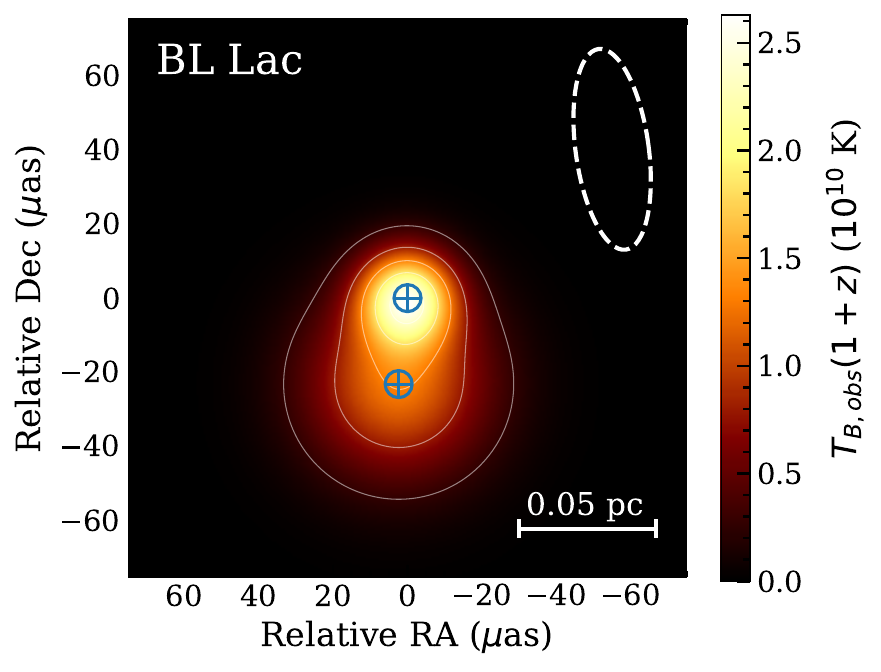}\hspace{0.09cm}
    \includegraphics[width=10.5cm]{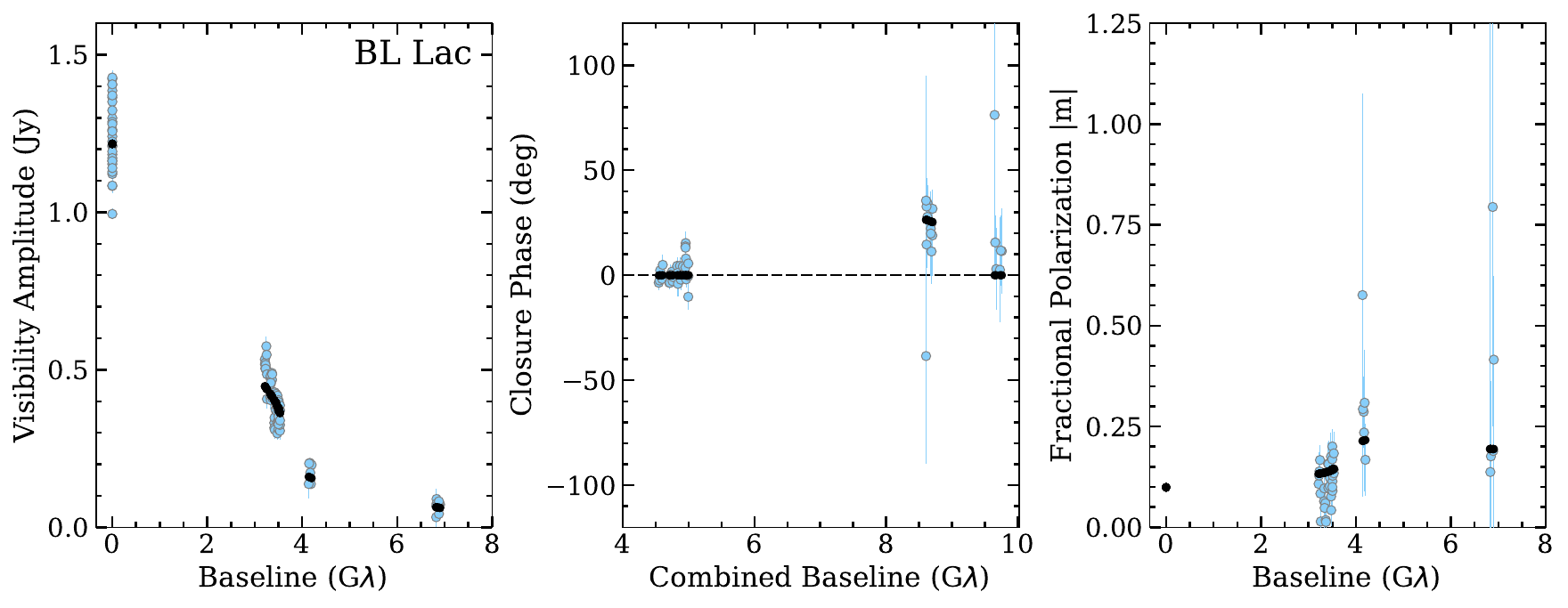}

    \includegraphics[width=5.25cm]{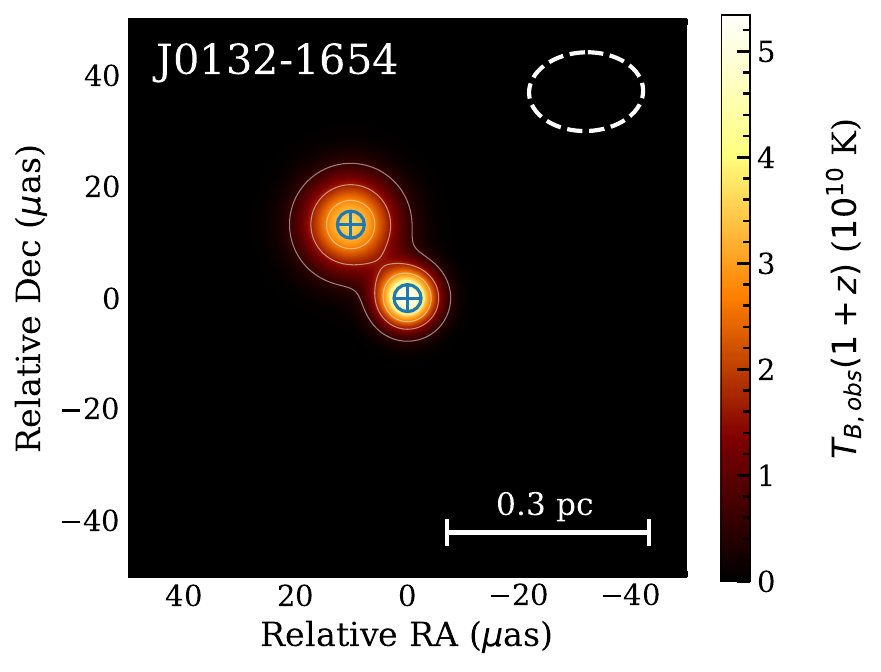}
    \includegraphics[width=10.5cm]{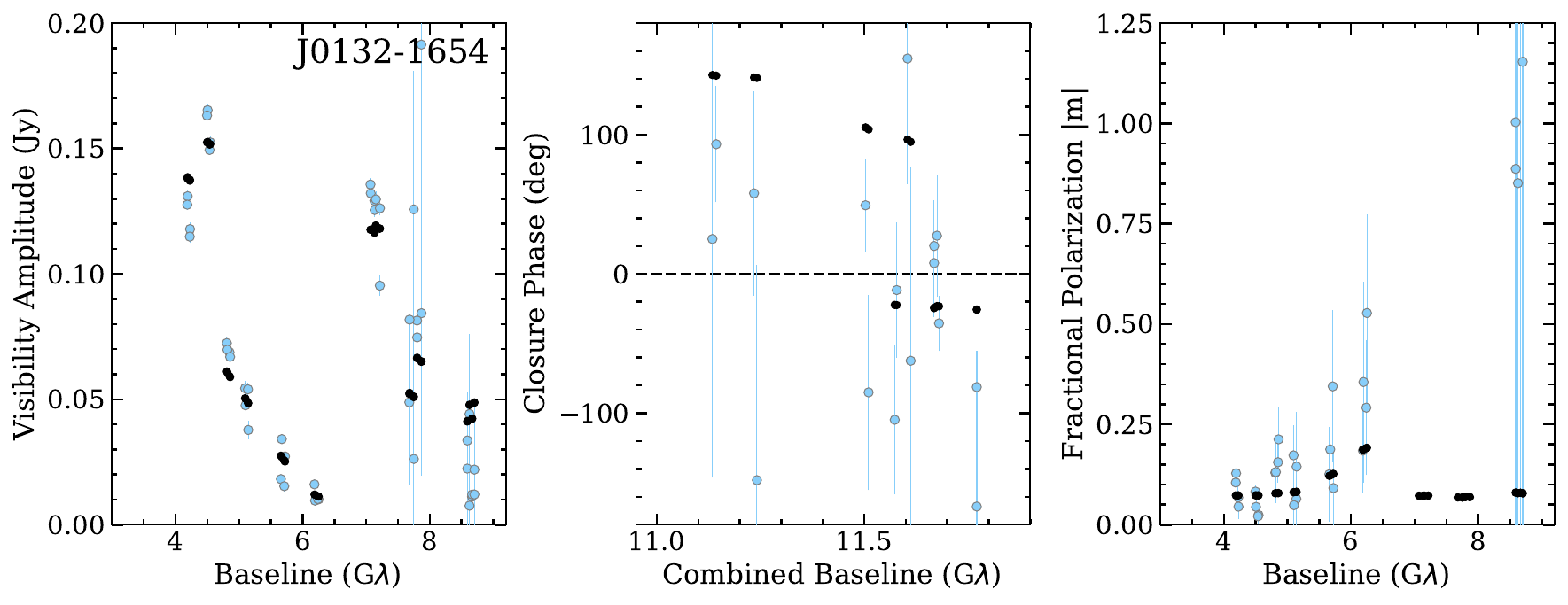}

    \includegraphics[width=5.25cm]{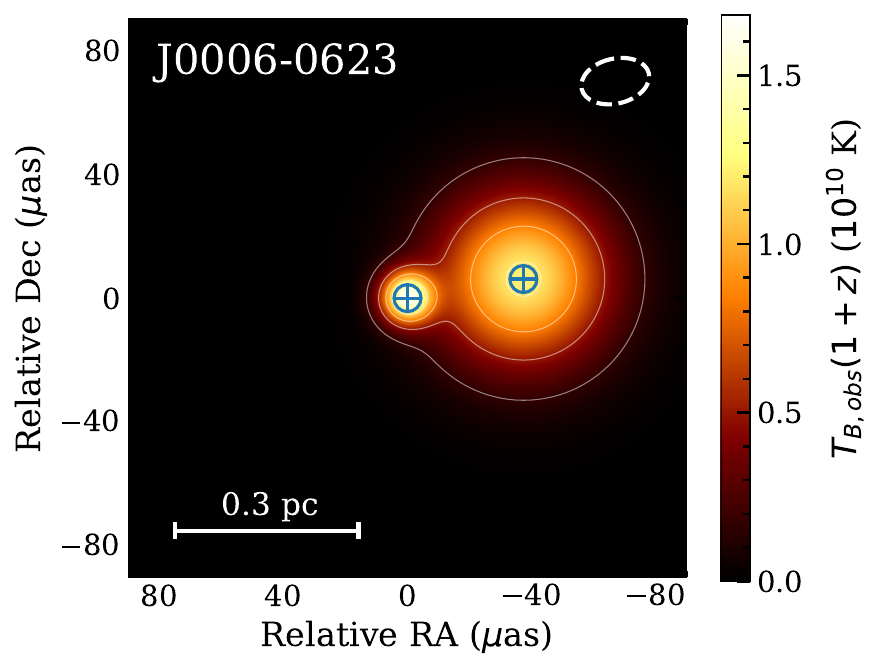}
   \includegraphics[width=10.5cm]{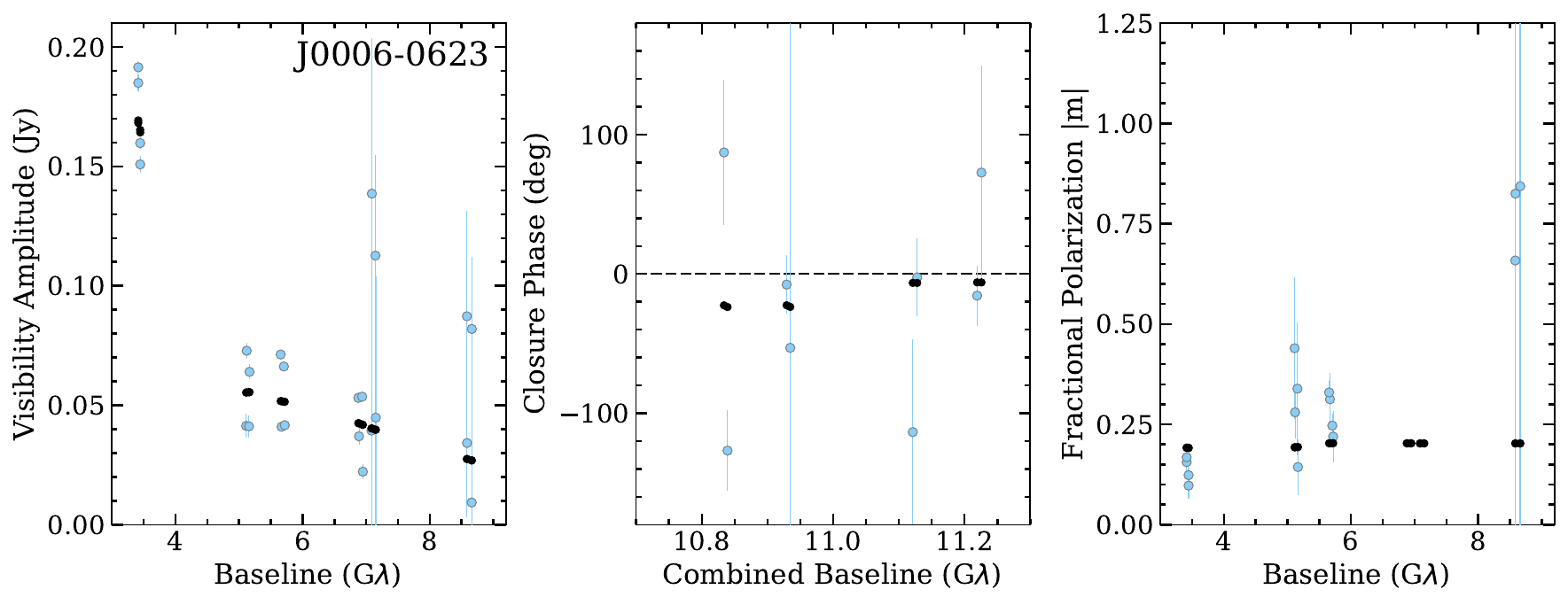}

   \vspace{-0.1cm}

 	\caption{Models of the EHT sources obtained through (polarized) circular Gaussian model fitting with at least two components. Blue crosses in the left column indicate positions of individual components. Contours represent 0.1, 0.3, 0.5, 0.7 and 0.9 of the peak total intensity. EHT observing beams are shown with dashed-line ellipses, to give a sense of the diffraction limited resolution supported by the data sets. We show consistency between models (black) and data (blue) for visibility amplitudes, closure phases, and absolute values of fractional visibility polarization.  Closure phases are shown as a function of combined baseline length, which corresponds to a quadrature sum of lengths of all baselines from a given triangle. \jref{The visibility domain fractional polarization may exceed unity for resolved sources, such as is the case for 1749+096.} The data shown contain detections only and no upper limits. }
  \label{fig:models_total}
\vspace{-0.2cm}
\end{figure*}

For five out of seven sources (1749+096, 1055+018, BL\,Lac, J0132--1654, J0006--0623) we modeled the morphology with two or three circular Gaussians, presented in Fig.~\ref{fig:models_total}. Low visibility amplitudes around $\sim$1\,G$\lambda$ for 1055+018 were identified as LMT miscalibration related to pointing issues \jref{(see also \citealt{M87p3} for a summary of issues with LMT in 2017). These points were downweighted for the amplitude fitting, but preserved for the closure phase fitting.} 
For the other two out of seven sources (CTA\,102 and 3C\,454.3), a~single circular Gaussian was sufficient to interpret the observations. \jref{We assume that the core components are nearly circularly symmetric.}

In the cases of BL\,Lac \citep{Casadio2021}, 1055+018 \citep{Weaver2022} and several other sources, the 230\,GHz model fit structure
is consistent with images obtained at lower frequencies. For other sources, the separation of scales 
between the highest angular resolution images available so far and our modeling results makes such a comparison 
difficult. In particular, we identify an east-west structure in 1749+096, well constrained by the data, 
but perpendicular to the jet observed at lower frequencies, extending in the north direction \citep{Weaver2022}. 
While in this paper we focus on the total intensity properties of the sources, we also obtained linear 
polarization results for the Gaussian components, reported in Appendix~\ref{app:modelfits} (Fig.~\ref{fig:models_info}) along with detailed parameters of the fitted Gaussian models (Table~\ref{tab:models}) and more 
technical comments regarding the model fitting procedures.

\begin{figure}
   \centering
\includegraphics[width=\columnwidth]{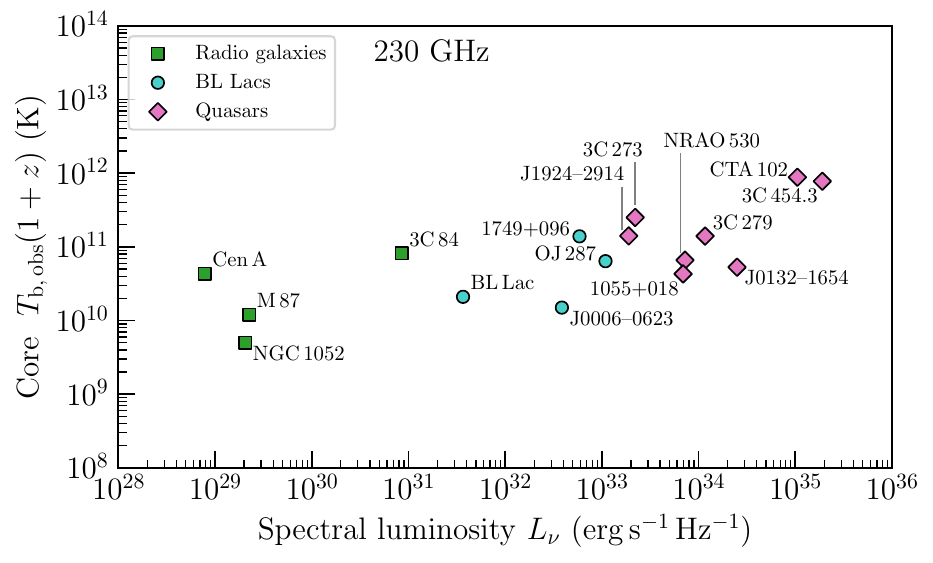}
   \caption{Core brightness temperature against core synchrotron spectral luminosity for the EHT 2017 sample of 16 AGN sources. The sources cluster into categories of radio galaxies, BL~Lacs (blazars), and quasars.}
   \label{fig:Tb_Lsyn}
\end{figure}

The EHT measurements of the 230\,GHz VLBI cores, discussed further in this work, can be represented on a plane of core brightness temperature (see Section~\ref{sec:Tb}) against isotropic spectral core luminosity $L_\nu \propto S_\nu D_L^2$ for the luminosity distance $D_L$, see Fig.~\ref{fig:Tb_Lsyn}. This representation emphasizes the differences between the types of observed sources within the inhomogeneous EHT data set, with radio galaxies at the lower luminosity and brightness temperature corner of the figure, and luminous, high brightness temperature quasars in the opposite corner. The figure does not show Sgr~A$^*$, with 230\,GHz spectral luminosity $L_v \sim 10^{23}$\,erg\,s$^{-1}$\,Hz$^{-1}$, over five orders of magnitude below the least luminous AGN in the sample.


\section{Measurements in the EHT$+$ data set} \label{sec:EHT+}

We consider sixteen AGN sources observed by the EHT, as summarized in Tables \ref{tab:source_list} and \ref{tab:sources}.      
For this set of sources, we additionally use measurements obtained at lower frequencies. We refer to this collection of measurements as the EHT+ data set. The EHT+ data set below 230\,GHz spans a range of frequencies from 2 to 86\,GHz, taken from a variety of surveys and monitoring efforts. At 1.66, 4.84, and 22.24\,GHz we used measurements from the RadioAstron space VLBI program (\citealt{Kardashev2013,Kovalev2020}, Kovalev et al., in prep.), comparing them to VLBA data at 2.3\,GHz \citep[single-epoch, ][]{Pushkarev2012} and 24\,GHz \citep[multi-epoch,][]{deWitt2023}, as well as measurements from the VSOP program at
5\,GHz \citep{Dodson2008}. Further VLBA data include 8\,GHz \citep{Pushkarev2012}, 
15\,GHz \citep[MOJAVE,][]{Homan2021},
and 43\,GHz \citep[VLBA-BU-BLAZAR,][]{Weaver2022}. At 86\,GHz, we used results obtained in a single-epoch GMVA AGN survey \citep{Lee2008}. In addition to these survey results, we used core 
brightness temperature measurements 
for M\,87 at 43\,GHz \citep{cheng2020} and Centaurus\,A at 8.4 and 22.3\,GHz \citep[TANAMI\footnote{Tracking active galactic nuclei with Austral milliarcsecond interferometry},][]{Muller2011}.
In order to increase the homogeneity of the data set and limit the impact of low resolution bias we exclude measurements obtained with arrays lacking long baselines, such as those procured using the Korean VLBI Network (KVN) for frequencies of 23, 43, 86, and 129\,GHz \citep{Lee2016_II}.

\begin{table*}[htp]
\caption{Flux densities, core sizes, and core brightness temperatures at 230\,GHz measured during the 2017 EHT observing campaign}
\vspace{-0.5cm}
\begin{center}
\begin{tabularx}{1.0\linewidth}{lccccccccc}
\hline\hline
Source & \multicolumn{5}{c}{ ------------------------ Flux Density (mJy) ------------------------} & Size  & $T_{\rm b, obs}(1+z)$ & $T_{\rm b,\,min}(1+z)$ & $T_{\rm b,\,lim} (1+z)$  \\
 & ALMA$^a$ & J-S & A-X & compact & core & $\upmu$as  & $10^{10}$\,K  & $10^{10}$\,K & $10^{10}$\,K \\
\hline

%
%

Sgr\,A$^*$  &     $2450 \pm 250 $  & $1867 \pm 312$ & $1920 \pm 140 $ & 2270 &  2270 & 53 & 1.4 & 1.1 & 1.2  \\

3C\,279   &     $8800\pm 800 $  & $7050\pm 600$ & $6510 \pm 300 $ & 7890 &  2500 & 25 & 14.0 & 7.9  & 20.9 \\

M\,87   &     $1310\pm 130$  & $980\pm133$ & $920 \pm 80$ &  610 & 610 & 42 & 1.2 & 0.2 & 0.2 \\

OJ\,287  &     $4270 \pm 430 $  & $3085\pm 1100$ & $2970 \pm 230$ & 3690 & 420 & 14 & 6.4  & 1.8 & 2.2    \\

Cen\,A  &     $5660 \pm 570 $  & $4816\pm 367$ & $4300\pm 120$ & 1750 & 1750 &  30 & 4.3 & 2.0 &  3.3  \\

J1924--2914 & $3200\pm320$ & $2472\pm345$ & $2550 \pm 170$ & 2470 & 500 & 11 & 14.1 & 19.7  & 78.9  \\

NRAO\,530   &     $1590\pm 160 $  & $1248\pm 141$ & $1140 \pm 60$ & 680 & 240  & 13.0 & 6.6 &  4.3 & 6.9  \\

3C\,273  &     $7560 \pm 760 $  & $6163\pm 700$ &  $5220 \pm 210 $ & 6000 & 610 & 8.1  & 25.1 & 9.1  & 38.8  \\

1749+096 & -- & $1806\pm 135$ & -- & 1600 & 670 &  12.9 & 12.6 & 17.4  & 73.9  \\

1055+018 & $3550\pm400$ & $2510 \pm 140$  & $2530 \pm 120 $ & 1240 & 315 & 18.0 & 4.3 & 3.4  & 14.4  \\

BL\,Lac & -- & $1241\pm 112$ & -- & 1220 & 1220 & 21.1 & 2.1  &  1.9 & 6.7  \\

3C\,84 & -- & $9895\pm 321$ & $7120 \pm 440 $& 1880 & 1040  & 17.5 & 8.2 &  6.8 &  29.0   \\

3C\,454.3 & -- & $9126\pm543$ & -- & 9040 & 9040 & 22.5 & 77.7 & 44.9  & 190.5  \\

CTA\,102 & -- & $5475\pm254$ & -- & 5400  & 5400 & 17.1 & 88.2 & 81.0 & 344.3 \\

J0132--1654 & $415\pm30$ & -- & -- & 212 & 80 & 8.5 & 5.3 & 2.4  & 3.2  \\

J0006--0623 & $1990\pm200$ & -- & -- & 930 &  91 & 13.9 & 1.5 &  2.4 & 3.9  \\

NGC\,1052    &     $400\pm50$ &  $350\pm 50 $ & -- &  $350 $ &  $350 $  & 43.0 & 0.5 & 0.1 & 0.2 \\

Cyg\,X--3\tablefootmark{b}    &     -- &  $1770\pm 349 $  &  --  & -- & -- & -- & -- & -- & --  \\

\hline

\label{tab:sources}
\end{tabularx}
\end{center}
\vspace{-0.7cm}
\tablefoot{\tablefoottext{a}{following \citet{Goddi2021}, corresponding to $\sim$1\,arcsec angular scale.} \tablefoottext{b}{Measurements on April 6 and 7, 2017 with the IRAM 30m telescope at 228.7\,GHz yielded respectively $S=(2.95\pm0.44)$\,Jy and $S=(2.96\pm0.26)$\,Jy \citep{Krichbaum2017}.} J-S: JCMT-SMA  baseline ($\sim$\,100\,k$\lambda$ or $\sim$1~arcsec); A-X: ALMA-APEX baseline ($\sim$\,1.5\,M$\lambda$ or $\sim$100~mas); compact: flux density constrained with long EHT baselines on a~sub-mas scale.}
\end{table*}

\subsection{Multi-frequency power law fits}
\label{subsection:PLmethod}

We aim to characterize the dependence of the measured quantities (VLBI core size, flux density, brightness temperature) on the frequency with a power law model. However, individual properties of sources differ, scaling the observables. \jref{These include}  their cosmological redshift, Doppler factor, intrinsic source power and distance, \jref{as well as their intrinsic variability}. Hence, the scaling parameter $b$ in a power law $b\nu^a$ is generally source-specific and the inhomogeneous data set may not be self-consistently fitted with a single power law. Thus, in this paper we treated the scaling $b$ as a source-specific nuisance parameter and characterized the slope $a$ for the EHT+ data set by studying 16 individual sources and subsequently aggregating the results. The parameter $b$ absorbs effects impacting the observables for the individual sources as a constant (frequency-independent) factor, such as cosmological redshift, constant Doppler factor, or intrinsic power. It preserves the relative impact of effects depending on frequency/location along the jet (acceleration, energy conversion), which we studied here.

To evaluate the characteristic power law slope in a frequency dependence of a given quantity for the entire EHT+ sample, we considered a set of $N=16$ slopes $a_i$, fitted separately for individual sources. We extracted their mean value $m$ and standard deviation $\sigma$, and used $m \pm \sigma/\sqrt{N}$ as our estimate of the characteristic power law slope in the population. In Appendix~\ref{Appendix:Doppler} we further discussed this choice and compare it with alternative approaches, such as directly fitting all measurements with a single power law, demonstrating robustness of the estimated slopes.

\subsection{Core size and flux density}

We estimated the VLBI core diameter $\theta$ and flux density $S_\nu$ using geometric Gaussian model fitting (see Section~\ref{subsec:modeling}), 
identifying the core with the brightest of the
fitted Gaussian components\jref{, that is, the component with the largest measured brightness temperature value}. Both core size and core flux density parameters are generally 
subject to significant systematic uncertainties, related to the sparse $(u,v)$--coverage. An interferometer is a spatial filter and the correlated flux density measured on the long baselines misses the resolved-out emission from structures larger than $\sim\!\lambda/{\rm BL}$, where BL is the baseline length.
This effect should not negatively affect the 
characterization of the compact cores with the EHT, but may be relevant for extremely long baselines and lower observing frequencies, as is the case for the RadioAstron observations.
Table~\ref{tab:sources} compares 230\,GHz flux densities measured by the connected ALMA array ($\sim$\,100-300\,k$\lambda$), the JCMT-SMA (J-S) baseline ($\sim$\,100\,k$\lambda$), as well as the ALMA-APEX (A-X) 
baseline ($\sim$\,1.5\,M$\lambda$). 
There are indications of losses in VLBI flux density measurements in comparison with connected-element ALMA interferometry for comparable baseline lengths (see Appendix~\ref{app:VLBIbias}). 
For the EHT data sets, ALMA flux densities can generally be considered to be the most reliable; 
thus, when available, they were used to calibrate the short VLBI baselines. This is the case in Table~\ref{tab:sources}, where we give VLBI measurements for short intrasite baselines without scaling them to ALMA measurements, but the compact VLBI and core flux densities follow subsequent rescaling for sources with sufficient ALMA-only data \citep[network calibration;][]{Blackburn2019}. Nonetheless, the additional flux density uncertainty of $\sim$\,20\% related to the ALMA-VLBI discrepancy (Appendix~\ref{app:VLBIbias}) is subdominant with respect to other systematics and does not significantly impact our results.

In contrast to the flux density, the estimate of the core size may indeed be affected by the limited instrumental resolution. The configuration of the 2017 EHT array yields extreme angular resolution, but suffers from generally sparse coverage and lack of baselines probing milliarcsecond scales. Therefore, the dynamic range of the reconstructions is often low, and extended components are resolved out. Hence, studying jets on milliarcsecond and larger angular scales with the 2017 EHT array is effectively impossible. The array is, however, well suited for studying dominant, bright and compact core components in AGN systems. Indeed, none of the considered data sets shows clear indication of compact unresolved structures, though some of the lower frequency measurements in the EHT+ data set do -- in those cases the core size estimates correspond to resolution-dependent upper limits.

The 15, 22, and 43\,GHz measurements used in this analysis were respectively obtained from year- or decade-long surveys as parts of the MOJAVE program, the K-band celestial reference frame survey, and the VLBA-BU-BLAZAR/BEAM-ME programs. At higher frequencies, such monitoring programs are not available; 86 and 230\,GHz measurements were collected in single-snapshot surveys. They may therefore not properly reflect the usual behavior of the individual variable sources. As an example, fluctuations of the 230\,GHz compact ring-like core of M87 between 0.5 and 1.0\,Jy were reported by \citet{Wielgus2020}. The intrinsic source variability potentially contributes to uncertainty, if, for instance, a source happened to be in a flaring state during the observation. In the case of RadioAstron data, the correlation between increased source activity and availability of detections was addressed by treating the obtained brightness temperatures as upper limits.

\begin{figure}[!htbp]
    \centering
    \includegraphics[width=\columnwidth]{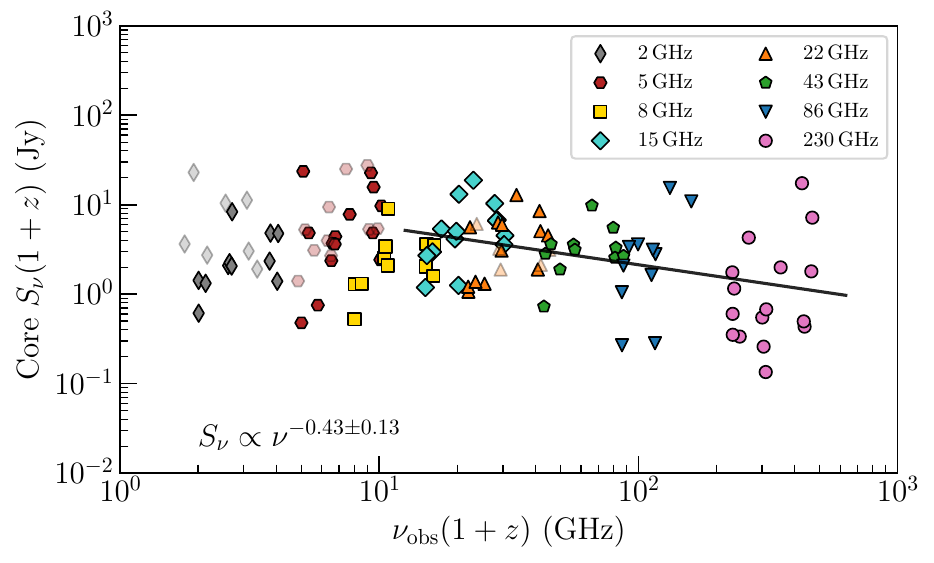}
    \includegraphics[width=\columnwidth]{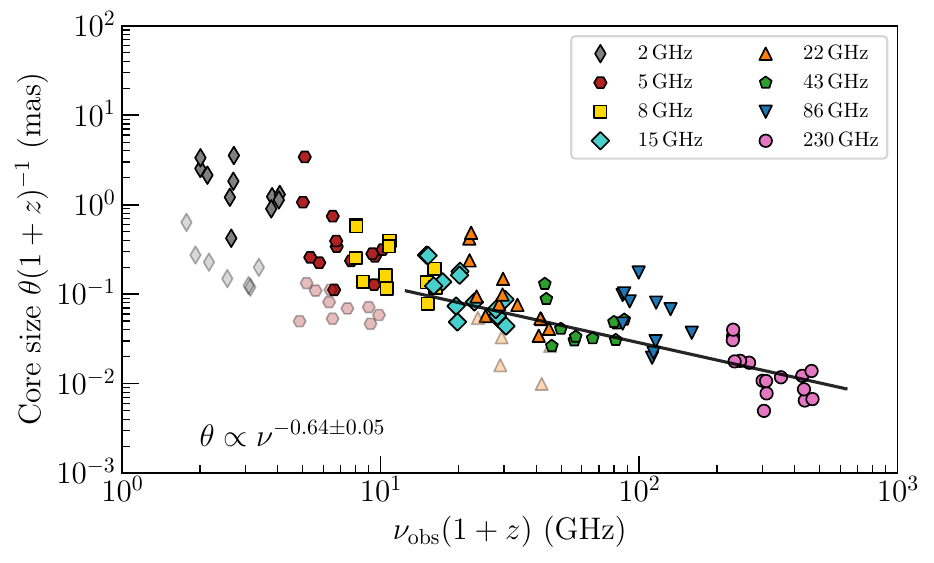}
    \includegraphics[width=\columnwidth]{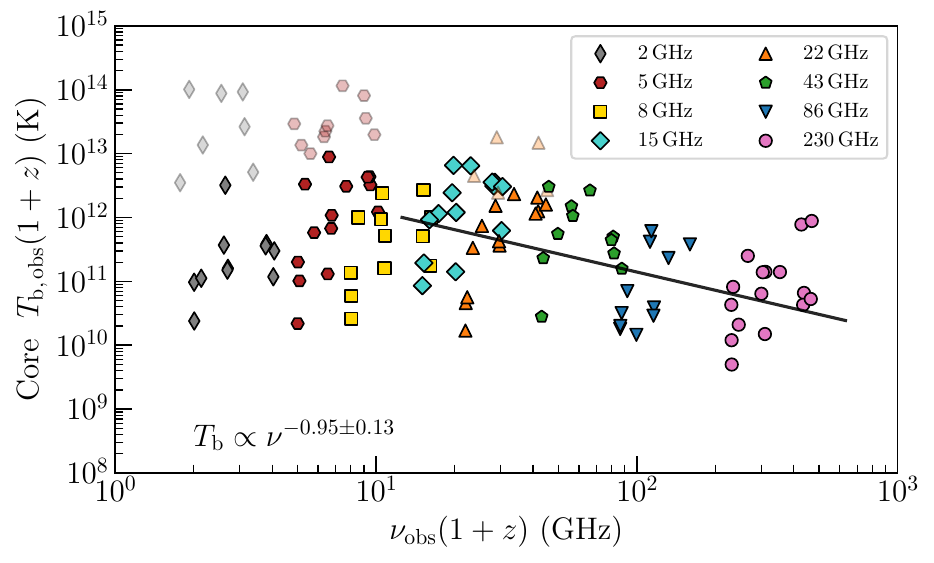}
    \caption{Measurements of the core flux density (top), size (middle), and brightness temperature (bottom) obtained using the EHT+ data set of AGN sources, as a function of frequency in the host frame. The 15--230\,GHz ground array data in each panel are approximated by power law fits (solid lines, \jref{obtained as detailed in Section \ref{subsection:PLmethod}}) and the fit results are annotated. The faded data points are the RadioAstron measurements, while regular data points at the same frequencies correspond to VLBA and VSOP measurements; \jref{for comparability, we color the 1.66\,GHz (L band) RadioAstron points the same as the 2\,GHz  (S band) VLBA points}.
    The slope of the core flux density is shallow, $S_\nu \propto \nu^{-0.4}$.
    The core size decreases with frequency (and, in turn, increases with the distance from the 
    central engine) as $\theta \propto \nu^{-0.6}$; the brightness 
    temperature decreases with frequency as $T_{\rm b} \propto \nu^{-1.0}$.}
    \label{fig:FluxSizeTb}
\end{figure}

Figure \ref{fig:FluxSizeTb} shows the flux density and size of the VLBI cores in the EHT+ data set, as well as the resulting brightness temperatures, against frequency measured in the host frame of the AGN (corrected for the cosmological redshift, hereafter referred to simply as "host frame"). For 2, 5, and 22\,GHz there are two sets of points, high angular resolution RadioAstron (semi-transparent markers) and lower resolution observations (solid-color markers)
\jref{. The systematic difference in the estimated core size is evident, with RadioAstron finding cores about an order of magnitude more compact. As a consequence, brightness temperatures inferred from RadioAstron observations sometimes approach $10^{14}$\,K, which is difficult to reconcile with the assumption of incoherent synchrotron emission from relativistic electrons as it would require untypically high Doppler boosting \citep{Kovalev2016}. Another effect possibly limiting the accuracy of the obtained measurements, particularly at lower frequencies, is related to blending between the core and the foreground jet components.}

In Fig.~\ref{fig:FluxSizeTb} we also present power law fits to ground array data from 15--230\,GHz, that is, excluding the lower frequency measurements suffering from the aforementioned systematic shortcomings. The calculation of the power law slope follows the methodology described in Section~\ref{subsection:PLmethod}. 
The spectra are almost flat (top panel), at most slightly decreasing with frequency as $S_\nu \sim \nu^{-0.4}$ for 15\,GHz and above, as expected from VLBI cores at high observing frequencies. The estimated core size decreases as $\theta\sim\nu^{-0.6}$ in the range of 15--230\,GHz. This decrease is less steep than the expected dependence dominated by the instrumental resolution effects $\sim\nu^{-1}$. The core size vs frequency panel of Fig.~\ref{fig:FluxSizeTb} indicates steepening of the slope at 2--8 GHz, in the region possibly more affected by the limited resolution \rev{and potentially also by scatter-broadening}. This further justifies only selecting 15--230\,GHz measurements for the power law fitting.

\subsection{Brightness temperature}
\label{sec:Tb}

The brightness temperature \tb is an important tool to probe the nature and launching mechanisms of astrophysical jets. In 
the emitter frame, the intrinsic brightness temperature $T_{\rm b,\,int}$ equals the equivalent black body temperature given the source surface brightness. It serves as a proxy of the temperature of electrons emitting the observed synchrotron radiation, enabling the 
characterization of the energy partition between plasma and the magnetic field, parametrised with the ratio of emitting particles energy to magnetic energy $\eta$ assuming self-absorbed synchrotron radiation \citep{Readhead1994,Homan2006}. 

The observed $T_{\rm b,obs}$ differs from the 
intrinsic value by a cosmological factor $(1+z)^{-1}$ (usually known well) and the Doppler factor $\delta$ depending on the bulk 
outflow velocity and the jet viewing angle (usually known poorly),
\begin{equation}
T_{\rm b,\,obs}=\frac{\delta}{1+z} \, T_{\rm b,\,int} = \frac{\delta}{1+z} \eta^{2/17} T_{\rm b,\, eq} = \frac{1}{1+z} T_{\rm b} \, ,
\label{eq:TB_maindef}
\end{equation}
\noindent
where $T_{\rm b,\,eq} \approx 5 \times 10^{10}$\,K is the equipartition brightness temperature \citep{Readhead1994}. 
Following Eq.~\ref{eq:TB_maindef}, in this work we denote the cosmological redshift-corrected brightness temperature (that 
is, measured in the frame of the host galaxy) simply with $T_{\rm b} \equiv T_{\rm b, obs} (1+z) = \delta T_{\rm b,int}$. For a circular Gaussian 
source model the peak brightness temperature in Kelvins can be calculated as
\begin{align}  
    T_{\rm b} &= 1.22 \times 10^{9} \left(\frac{S_\nu}{\rm mJy}\right) \left(\frac{\nu}{\rm GHz}\right)^{-2}  \left(\frac{\theta}{\rm mas}\right)^{-2}\!(1+z) \nonumber \\
    &= T_{\rm b, obs}(1+z) \, ,
\end{align}
where $S_\nu$ is the measured correlated flux density at the observing frequency $\nu$, and $\theta$ is the angular size of 
the source, defined as the full width at half maximum (FWHM) of the fitted Gaussian. By using a simple Gaussian model, we neglect the transverse structure of the jet. The frequency in the host frame is larger than the one observed by a factor $(1+z)$, but in the comoving frame of the emitting plasma it is reduced by a factor of $\delta$, typically exceeding unity for jet sources. 

Since modeling the source structure using sparse VLBI data is subject to large systematic uncertainties, model-agnostic estimates based solely on visibility measurements provide potentially useful limits on the brightness temperature \citep{Lobanov2015}: 
\begin{align}
T_{\rm b,\, min} &\approx 3.09 \left(\frac{\rm BL}{\rm km}\right)^2 \left(\frac{V_q}{\rm mJy}\right)\, ,\label{eq:Tbmin}\\
T_{\rm b,\, lim} &\approx 1.14 \left(\frac{\rm BL}{\rm km}\right)^2 \left(\frac{V_q+\sigma_q}{\rm mJy}\right) \left(\ln\frac{V_q+\sigma_q}{V_q}\right)^{-1}\, ,\label{eq:Tbmax}
\end{align} 
with maximum baseline length BL, corresponding visibility amplitude $V_q$ and its uncertainty $\sigma_q$. The values of BL are reported in Table~\ref{tab:source_list};
core component flux density, size, and peak brightness temperature of the source model $T_{\rm b,obs}(1+z)$, as well as the visibility-only brightness temperature estimates can be found in Table~\ref{tab:sources}. The latter indicate broad consistency with the brightness temperatures obtained based on Gaussian component modeling. 

In the case of sources with relatively good $(u,v)$-coverage, with detailed analyses described in dedicated papers \citep[e.g.,][Gomez et al. in prep.]{Janssen2021}, we report core parameters following the imaging results presented therein, without resorting to approximated geometric modeling with Gaussian components.

The resulting measurements of brightness temperature $T_{\rm b}$ are shown in the bottom panel of 
Fig.~\ref{fig:FluxSizeTb}. The systematic difference between measurements with long RadioAstron baselines (semi-transparent markers at 2, 5, and 22 
GHz) and the other observations at the same frequencies obtained with the VLBA and as part of the VSOP program 
are clearly visible. A power law fit to the data at frequencies of 15 GHz and larger indicates a slope with an index of $-0.95\pm0.13$, fitted with the methodology described in Section~\ref{subsection:PLmethod}; see also the discussion in Appendix~\ref{Appendix:Doppler}.


\section{Modeled quantities}
\label{sec:modeled_quantities}
\subsection{Distance from the VLBI core to the black hole}\label{sec:dist_core}

We adopt the framework for relativistic jets established by \citet{Blandford1979} and \citet{Konigl1981}, assuming a supersonic, conical 
jet with an opening angle $\phi_{\rm o}$, and a viewing angle~$\iota$. We 
refer to this setup as the BK model. The jet bulk Lorentz factor $\gamma_{\rm j}$ is constant in this framework, and the jet magnetic field $B \propto r^{-m}$ and particle density $N \propto r^{-n}$ are described as functions of the 
distance $r$ from the jet origin.

Following \citet{Lee2016_I}, we employ a measure for the distance of the observed VLBI core to the true central engine under the assumption of equipartition between the particles in the jet and the magnetic field. 
The VLBI core is defined as the region where the optical depth reaches unity. Then, the physical distance (measured along the jet) of the VLBI core to the true central engine is \citep{Lobanov1998}: 
\begin{equation}\label{eq:dist_1}
r =  \left( \frac{B_1^{k_{\rm b}}}{\nu(1+z)}\left[6.2\times10^{18} C_2(\alpha) \delta^\epsilon\phi_{\rm o} N_1\right]^{1/(\epsilon+1)}\right)^{1/k_{\rm r}} {\rm pc},
\end{equation}
where $B_1$ \jref{and $N_1$ are, respectively, the magnetic field strength and
electron number density} at $r_1$ = 1\,pc distance from the jet origin, $\delta$ is the jet Doppler factor $\delta = (1 - \beta \cos \iota)^{-1} \gamma^{-1}$, $C_2(-0.5)=8.4\times10^{10}$ cgs \citep{Pacholczyk1970,Konigl1981} and
\begin{align}
    k_{\rm r} &= \left[(3-2\alpha)m+2n-2\right]/(5-2\alpha), \label {eq:kr} \\
    k_{\rm b} &= (3-2\alpha)/(5-2\alpha),\\
    \epsilon &= 3/2 - \alpha.
\end{align}
In this work we do not attempt to use Lorentz factors, Doppler factors, and viewing angles measured for individual sources. Instead, 
following previous analyses, we assume a characteristic bulk Lorentz factor $\gamma_{\rm j}=10$ for the entire sample and $N_1=5\times10^3$\,cm$^{-3}$ at a distance of $r_1$ = 1\,pc from the black hole \citep{Lee2016_I}.
For the intrinsic and observed opening angles, we set ${\phi=0.01\,{\rm rad}\approx0.6^\degree}$ and $\phi_{\rm o}=\phi \csc\iota$ with the viewing angle $\iota=0.1$\,rad, resulting in $\delta \approx 10$. Furthermore, following \citet{Konigl1981} and \citet{Lobanov1998} we 
assume energy equipartition and adopt $m=1$, $n=2$, $k_{\rm r}=1$, $k_{\rm b}=2/3$, and ${\epsilon=2}$, with $\alpha = -0.5$ ($S_\nu \propto \nu^{\alpha}$). The magnetic field strength $B_1$ at 1\,pc can be expressed through the 
total synchrotron luminosity $L_{\rm syn}$, following \citet{Blandford1979}:
\begin{equation}
    L_{\rm syn} = 4 \pi D_{\rm L}^2 S_{\rm int} \propto \gamma_{\rm j}^2 \beta_{\rm j}c B_1^2 r_1^{2m} \phi^2 ,
\end{equation}
where $D_{\rm L}$ is the luminosity distance to the source and $S_{\rm int}$ is the integrated, redshift corrected observed synchrotron flux density, integrated in the host frame frequency range between 1\,GHz and 700\,GHz by fitting a power law in $\nu$ to measured $S_\nu$ of each individual object. Equation \ref{eq:dist_1}
then takes a form 
\begin{equation}
r = \left(\frac{ K L_{\rm syn}^{k_{\rm b}/2}}{\nu(1+z)}\right)^{1/k_{\rm r}} = \frac{K}{\nu (1+z)} L_{\rm syn}^{1/3} \ \  {\rm pc},
\label{eq:r_pc}
\end{equation}
with a constant $K$ incorporating the assumed BK model parameters. In order to correct $L_{\rm syn}$ for the Doppler effect, the right hand side of the Eq.~\ref{eq:r_pc} would be scaled by $\sim\delta^{-1}$, where the exact power depend on detailed physical assumptions \citep{Ghisellini1993}. The other Doppler factor present in Eq.~\ref{eq:dist_1} was absorbed into the $K$ factor in Eq.~\ref{eq:r_pc} and for the assumed parameters corresponds to $\delta^{2/3}$, so the overall dependence of the radius estimate on the Doppler factor is shallow $\delta^{-1/3}$. In previous 
studies, the model described above was applied to 
measurements made at frequencies up to 86\,GHz \citep{Lee2016_I,Nair2019}. 

\begin{figure}
    \centering
    \includegraphics[width=\columnwidth]{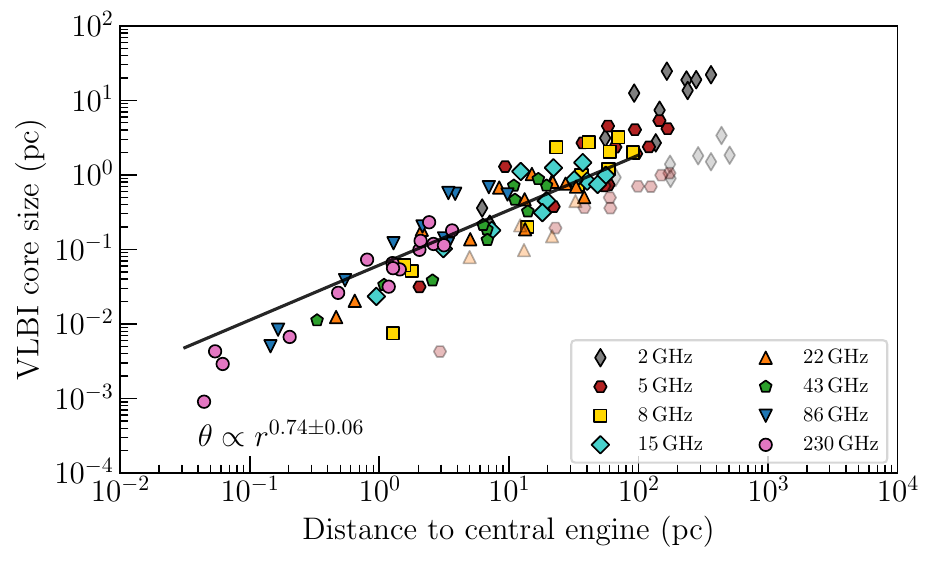}
    \includegraphics[width=\columnwidth]{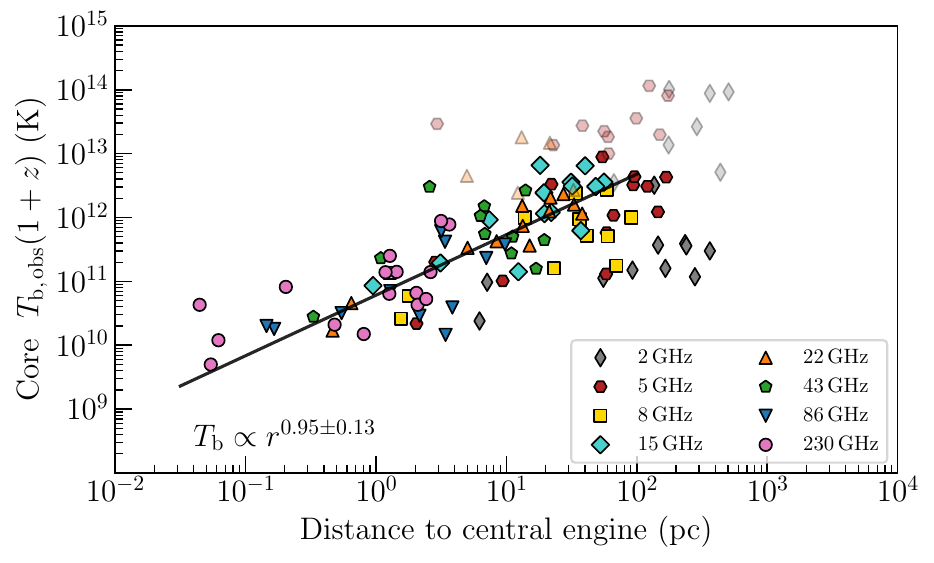}
    \includegraphics[width=\columnwidth]{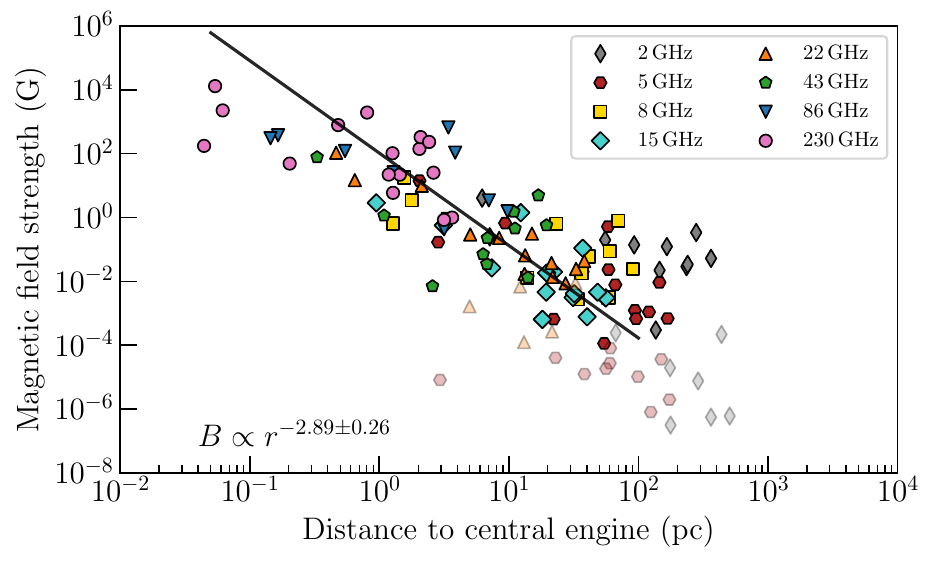}

    \caption{Core size (top), brightness 
    temperature (middle) and magnetic field estimate (bottom)
    obtained from the EHT+ data set as a function of the distance to the central engine. 
    The 15--230\,GHz data in each panel are approximated by power law fits (solid lines, \jref{obtained as detailed in Section \ref{subsection:PLmethod}}) and the fit results are annotated. The faded data points at are the RadioAstron measurements, while the filled markers correspond to VLBA observations. \jref{For comparability, we color the 1.66\,GHz (L band) RadioAstron points the same as the 2\,GHz  (S band) VLBA points}. For a $M_{\sbullet[1.35]} = 10^8 M_{\odot}$ black hole 1\,pc $ = 2 \times 10^5 r_{\rm g}$ with gravitational radius $r_{\rm g} = GM_{\sbullet[1.35]}/c^2$.}
    \label{fig:resultsBK}
\end{figure}

The choice of particular fixed BK model jet parameters for an inhomogeneous sample of sources like EHT+ is justified by the fact that the sample is dominated by quasars and BL Lac objects. Furthermore, we are mostly interested in the power law dependence. Essentially, with the methodology described in Section~\ref{subsection:PLmethod}, the only BK model information impacting the power law index fits shown in Fig.~\ref{fig:resultsBK} is that $r \sim \nu^{-1/k_r}$. We comment further on the impact of source-specific corrections in Appendix~\ref{Appendix:Doppler}, where we incorporate Doppler factor corrections following estimates given in Table~\ref{tab:BHmasses}.

\subsection{Magnetic field strength}\label{sec:magfield}

The magnetic field strength of a synchrotron self-absorbed core can be roughly estimated as \citep[\rev{e.g., Section 5.3 of}][]{Condon16}
\begin{equation} \label{eq:magfield_condon}
    B \approx 1.4\times 10^{21} \left(\frac{\nu_{\rm obs}(1+z)}{\rm GHz}\right)\left(\frac{T_{\rm b,\,obs}(1+z)}{\rm K}\right)^{-2} \, {\rm G},
 \end{equation}
where we do not attempt to correct for the Doppler factor, which would increase $B$ in the emitter's frame by a factor $\delta$. While this 
estimate is independent of the BK jet model assumptions, it incorporates a 
very simplistic model for the emission spectrum. 
We found that Eq.~\ref{eq:magfield_condon}, while having the same functional dependence of $B \propto \nu T_{\rm b}^{-2}$, results in a magnetic field $\sim$25 times 
stronger compared to the $B_{\rm SSA}$ estimator of \citet{Marscher1983} for 
identical input; the latter, however, is only applicable at the synchrotron 
turnover frequency. \rev{Hence, we expect a systematic upward bias of $B$. Nonetheless, the relative differences and the slopes remain useful for interpretation, provided that the VLBI cores do not become optically thin at the higher observing frequencies.}



\section{Results and discussion}

The observations collected within the EHT+ data set are presented in 
Fig.~\ref{fig:FluxSizeTb}. In the framework of the BK jet model, we expect the intrinsic brightness temperature 
$T_{\rm b, \, int}$ to not exceed the equipartition limit ${T_{\rm b,\,eq}\leq 10^{11}}$\,K 
\citep{Readhead1994,Lahteenmaki1999_II,Singal2009} and, more fundamentally, the inverse Compton limit 
${T_{\rm b,\,IC}\sim 5\times10^{11}}$\,K \citep{Kellermann1969,Nair2019}. 
Observed brightness temperatures in excess of these limits may be caused by large Doppler factors as ${T_{\rm b, \, obs} (1+z) = \delta_{\rm eq} T_{\rm b, \, eq}}$. The equipartition Doppler factors 
necessary to fulfill this condition are $\delta_{\rm eq}<10$ in the EHT sample at 230\,GHz. Hence, from measured values of $T_{\rm b}$ alone, all sources are consistent with the equipartition limit without invoking unreasonably high Doppler factors at high observational frequencies. When interpreting brightness temperature measurements, we additionally make the crucial assumption that we observe self-absorbed, optically thick cores, and that they do not become fully optically thin at high observing frequencies.

In some cases, we observed 230\,GHz brightness temperatures significantly below $T_{\rm b,\,eq}$, which are better explained by a~magnetically dominated inner jet, in which case the observed brightness temperature is reduced by $\eta^{2/17} < 1$, following Eq.~\ref{eq:TB_maindef}. On the other hand, very large brightness temperatures obtained by RadioAstron 
at 1.7 and 5\,GHz, reaching $10^{14}$\,K, are difficult to reconcile with the assumption of equipartition, requiring unrealistically 
large equipartition Doppler factors. 
Instead, a more complicated core geometry or scattering sub-structure may play a role in driving up the brightness temperature \citep[e.g.,][]{2015ApJ...805..180J,2016ApJ...820L..10J}. 
The RadioAstron measurements were excluded from the power law fitting, which was limited to 15--230\,GHz ground array data.

\subsection{Tweaking the Blandford-K\"onigl model parameters}


The new observations at 230\,GHz confirm that the brightness temperature $T_{\rm b}$ increases with the distance from the central engine, that is, it becomes larger for lower observing frequencies.
Indications of such a 
trend were previously found by \citet{Lee2008} and \citet{Nair2019PhD}. 
While quantifying these trends is hindered by significant systematic 
uncertainties, in Section~\ref{sec:EHT+} we reported the core size and brightness temperature dependence on the observing frequency $\theta \propto \nu^{-0.6}$ and  $T_{\rm b} \propto \nu^{-1.0}$.

To estimate the radial distance from the black hole to the VLBI core, we set up a jet model using various assumptions, 
as described in Section~\ref{sec:modeled_quantities}.
For a conical BK jet, \rev{the core diameter  
scales with frequency as $\theta \propto r \propto \nu^{-1/k_r}$,} where $k_r = 1$ in the case of equipartition. This immediate tension with 
the $\theta \propto \nu^{-0.6}$ dependence observed in Fig.~\ref{fig:FluxSizeTb} could be alleviated by a larger value of $k_r$.
Following 
Eq.~\ref{eq:kr}, $k_r$ depends on the assumed radial density and magnetic field strength profiles; a larger $k_r$ corresponds to a faster decay of these quantities with radius. 
In our example, changing $B \propto r^{-1}$ to $B \propto 
r^{-2}$ would be enough to reconcile the conical BK model with the observed relationship between core size and frequency. A steep decrease of the modeled magnetic field strength with radius is also found under the BK assumptions, see the bottom panel of Fig.~\ref{fig:resultsBK} and Section~\ref{sec:mf_discussion}.

Alternatively, the mismatch between our measurements and the brightness temperature, jet diameter and magnetic field strength radial dependence predicted by the BK model may be due to a transition from a parabolic to a conical jet geometry.
There is ample observational evidence for a parabolic geometry of the jet base \citep[e.g.,][]{Asada2012,Hada2016,Hada2018,Okino2022,Ricci2022}. Calculating the distances to 
the central engine 
using the BK model (Section~\ref{sec:modeled_quantities} and Fig.~\ref{fig:resultsBK}), we find $\theta \propto r^{0.7}$, which deviates from the expectations for a conical jet 
($\theta \propto r^{1.0}$) in the direction of a more parabolic structure ($\theta \propto r^{0.5})$. However, this predominantly shows the inconsistency of the canonical BK model with the data.

The brightness temperature is a derivative quantity of measured core size and flux density, 
$T_{\rm b} \propto S_\nu\nu^{-2} \theta^{-2}$. 
For the canonical BK jet, $S_\nu$ is flat and $\theta \propto \nu^{-1}$, resulting in a flat $T_{\rm b}$. 
From the EHT+ measurements we found a mildly negative slope of $S_\nu$, and the slope of $\theta(\nu)$ is shallower than 
the BK prediction, adding up to the observed $T_{\rm b} \propto \nu^{-1.0}$ dependence. While 
adjusting $k_r$, as discussed above, would take care of the impact of the core size trend on $T_{\rm b}$, it would not 
address the impact of the flux density trend.

\subsection{Doppler factor evolution and energy conversion}


Alternatively, the assumptions of a constant jet velocity or energy partition factor may need to be abandoned, as ${T_{\rm b} \propto \delta \eta^{2/17}}$. With the power law slopes found in the bottom panel of Fig.~\ref{fig:FluxSizeTb} we obtain
\begin{equation}
T_{\rm b, eq} = T_{\rm b} \delta^{-1} \eta^{-2/17} \propto \nu_{\rm int}^{-0.95}\delta^{-1.95} \eta^{-2/17} \ ,
\end{equation}
where $\nu_{\rm int} = \nu_{\rm obs}(1+z)/ \delta$ is the frequency in the emitter's frame. 
If we assume $\eta = \rm{const.}$, we find a constant intrinsic brightness temperature for a physically reasonable $\delta \propto \nu_{\rm int}^{-0.5} \propto r^{0.5}$ for $k_r = 1$. Hence, the Doppler factor grows more rapidly in the region close to the black hole. Allowing for $\eta$ to increase with radius, as magnetic energy is transferred to particles, adds another degree of freedom and will generally decrease the slope of $\delta(r)$. If we additionally require a flat spectrum measured in the jet frame, given the observed $S_\nu$ slope (top panel of Fig.~\ref{fig:FluxSizeTb}), we arrive at $\delta \propto r^{-0.3}$ and $\eta \propto r^{-3.1}$. These findings are a direct consequence of the observations and are independent of the BK model assumptions other than the choice of $k_r$ with $\nu \propto r^{-k_r}$. 

The growth of the Doppler factor with radius is a well motivated conclusion
in the context of compact-scale AGN jets, since the bulk acceleration of the outflow must take place somewhere between the black hole and the parsec scales.
A model in which both $\delta$ and $\eta$ grow with radial 
distance from the black hole is consistent with a magnetically accelerated jet \citep{vlahakis2004}, 
transitioning from the magnetically (or Poynting) dominated innermost region to energy equipartition (or dominance of particle kinetic energy) further away. A different 
physical scenario has been proposed by \citet{Melia1989} and \citet{Marscher1995}, where an electron-positron jet is accelerated to ultra-relativistic energies at compact scales and subsequently decelerated through inverse-Compton scattering with external photons. In the 
process, high-energy emission in X-ray and $\gamma$-ray bands is produced, and the jet becomes progressively brighter in radio band further away from the 
black hole. This scenario was discussed in the context of the brightness temperature statistics by \citet{Lee2014}. We consider this model to be less physically plausible \jref{\citep[see, e.\,g.,][on the role of radiation drag for jet deceleration]{Sikora1996}.}

Another caveat is the possible impact of a change in viewing angle $\iota$ in a bending jet on the radial profile of the Doppler factor $\delta$. Observations 
confirm the curved structure of some of the compact-scale jets in this work \citep[e.\,g.,][]{Issaoun2022,Jorstad2023}, further increasing the spread of 
brightness temperatures measured in the EHT+ sample with the viewing angle varying between the observing frequencies. If a certain source was bright at low frequencies due to a favorable viewing angle, it would show a lower 
core $T_{\rm b}$ than expected at higher frequencies, given a bend away from the favorable inclination in the more inner part of the jet. Moreover, an 
acceleration to speeds above $\beta = v/c = \cos \iota$ (or $\gamma > \csc \iota$) would decrease the observed $T_{\rm b}$ again, as the radiation becomes 
increasingly beamed along the jet axis, away from the observer at inclination~$\iota$. At large angular scales, we expect jets to be better described by the 
BK model, with a flat core spectrum $S_\nu (\nu)$, a continued high frequency trend in $\theta(\nu)$, and a flattening of $T_{\rm b} (\nu)$. This seems to be 
the case for the 2 and 5\,GHz observations shown in Fig.~\ref{fig:FluxSizeTb}, although the conclusions are uncertain given the large spread and the small size of the sample.

\subsection{Magnetic fields}
\label{sec:mf_discussion}

The bottom panel of Fig.~\ref{fig:resultsBK} shows the magnetic field estimates against the distance to the black hole, calculated with the BK model. For a canonical BK model (i.\,e., flat $T_{\rm b}$), as described in Section~\ref{sec:modeled_quantities}, Eq.~\ref{eq:magfield_condon} gives $B \propto \nu \propto r^{-k_r}$, self-consistent assuming $m = k_r = 1$. However, since we measure $T_{\rm b} \propto \nu^{-1}$, Eq.~\ref{eq:magfield_condon} gives $B \propto \nu^{3} \propto r^{ - 3 k_r}$, consistent with the result from fitting the data points in the bottom panel of Fig.~\ref{fig:resultsBK}. However, $3 k_r = m$ would imply $m+ n = 1$, requiring very shallow dependence of gas density and magnetic field strength with radius. These observations are thus in tension with the BK model. Correcting for the Doppler factor in Eqs.~\ref{eq:r_pc}--\ref{eq:magfield_condon} would allow to mitigate the steepness of the radial dependence of $B$.
 

Magnetic 
field estimates for the most compact scales reach ${B\sim10^3\,{\rm G}}$, which is consistent with some predictions for magnetized accretion disks 
\citep{Field1993}. In the special case of M\,87, where the central engine can be resolved, a~correction for the over-estimation of the magnetic field (see Section~\ref{sec:magfield}) would bring down the obtained field strength to a value comparable to estimates made by the EHT in the previous studies \citep{M87p8}. Across the EHT+ sample the field strength decreases by about seven orders of magnitude towards the largest probed scales of $\sim\!10^6 r_{\rm g}$ ($\sim\!10\,{\rm pc}$). At distances larger than one parsec estimated fields of ${B\sim10^{-4}\,{\rm G}}$ become comparable to the $\upmu$G-field of the ambient medium  
\citep{McKee2007,Beck2015}. 
In the BK jet framework, the magnetic field components perpendicular and parallel to the jet axis behave as ${B_{\bot}\propto r^{-1}}$ 
\citep{Blandford1979} and ${B_{\parallel}\propto r^{-2}}$ \citep{Konigl1981}, respectively. 
We find that the $B(r)$ slope is steeper 
than $-1$, consistent with the steeper $B(r)$ slope inferred from the observed core size dependence on frequency. This supports the presence of poloidal (jet-parallel) magnetic fields in the inner jet regions, possibly forming a mixed helical geometry \citep{Gabuzda2017}. A steep decrease of the magnetic field 
strength with radius may also be indicative of an efficient conversion of magnetic energy into kinetic energy of particles through, for example, magnetic reconnection.
The estimated slopes for some of the sources inspected individually are steeper than $-3$, following the decrease of the observed brightness temperature with frequency. This may be a consequence of rapid 
acceleration and a related radial increase of Doppler factor, unaccounted for in the BK model. \rev{Given the dependencies of $B_\parallel$ and $B_\bot$, a shallower slope in between $-2$ and $-1$ could be interpreted as a helical, but coherent field; the steep measured slope could hence indicate a loss of magnetic field coherence or strong dissipation through magnetic reconnection at larger distances. A steep slope of $B(r)$ could also indicate a~decrease of the optical depth at high observing frequencies, biasing the magnetic field estimates upward in the more compact regions. }


\section{Summary and conclusions}

In this work, we presented an analysis of the full EHT 2017 observational data set: the first 230\,GHz 
VLBI campaign of this magnitude. We compiled the EHT+ sample of sixteen AGN sources observed by the EHT, along with 
their VLBI observations available at lower frequencies (2--86\,GHz). For seven of these AGN sources we presented visibility domain modeling of the EHT data; the analyses of the remaining nine sources were given in separate papers. We first studied the change of the VLBI core flux density, size, and brightness temperature as a function of frequency in the EHT+ data set. Despite large scatter in the measurements, related to individual source properties, the joint 
analysis reveals a shallow dependence of the core size on frequency $\theta \propto \nu^{-0.6}$ and a systematic decrease of the brightness temperature with frequency $T_{\rm b} \propto 
\nu^{-1.0}$, indicating an increase of brightness temperature with the distance from the AGN central 
engine. These findings are qualitatively consistent with previous studies using lower and fewer observing frequencies. 


We demonstrated that properties of AGN jet sources constrained by the VLBI observations at 15-230\,GHz are incompatible with the
standard BK model of a conical jet with constant Lorentz factor and energy partition. 
Discussing the impact of variations of the BK model parameters and the jet collimation profile led us to the conclusion that either a bulk 
acceleration of the jet (an increase of the Doppler factor with the jet radius), or a transfer of energy from the magnetic field to 
the emitting particles is required to interpret the data. 

Both effects may occur simultaneously, and both are 
expected to play a role in compact scale jets based on theoretical models. A radial dependence of the Doppler factor $\delta \propto r^{0.5}$(or a slightly more shallow one, in the case of a radially evolving energy partition), could explain the observations. Our findings are consistent with these effects 
occurring gradually across the innermost parsec of the jet, or within $\sim\!10^5\,r_{\rm g}$ from the central black hole, with most of the Doppler factor increase occurring very close to the central engine.

Additionally, using the BK model, we estimated a steep decrease of magnetic field with radius $B \propto r^{-3}$, which is in tension with the underlying assumptions. The steepness of the slope may be reduced by incorporating a radially increasing Doppler factor, once again hinting at bulk acceleration of the jet. Alternatively, a strong dissipation of the magnetic energy may be taking place in the compact region of the AGN jets.

Subsequent EHT campaigns will deliver 230\,GHz VLBI measurements for a larger number of objects, increasing our EHT+ 
sample size and thus its statistical robustness. With more high quality data it will become feasible to apply (possibly frequency dependent) Doppler corrections to individual sources. 
Further, studying jet kinematics on EHT scales through tracking 
of individual moving features and comparing these results with lower frequency VLBI data could 
conclusively demonstrate the radial profile of jet acceleration, breaking degeneracies in our theoretical models. Finally, an extension of VLBI 
capabilities to 345\,GHz, which is already in the process of being implemented within the EHT, will provide insight 
on AGN jets on even more compact scales.

\section*{Data availability}
\label{sec:data_avail}

A table compiling measured VLBI core flux densities, FWHM sizes and brightness temperatures, as well as the derived distances to the black hole and magnetic field strengths is available in electronic form at the CDS via anonymous ftp to \url{cdsarc.u-strasbg.fr} (130.79.128.5) or via \url{http://cdsweb.u-strasbg.fr/cgi-bin/qcat?J/A+A/}.




\bibliographystyle{aa}
\bibliography{bibliography}




\appendix 

\FloatBarrier

\onecolumn

\section{Models of the EHT sources}\label{app:modelfits}

\begin{figure*}[h!]
	\centering
    \includegraphics[height=3.1cm,trim={0 0 1.1cm 0},clip]{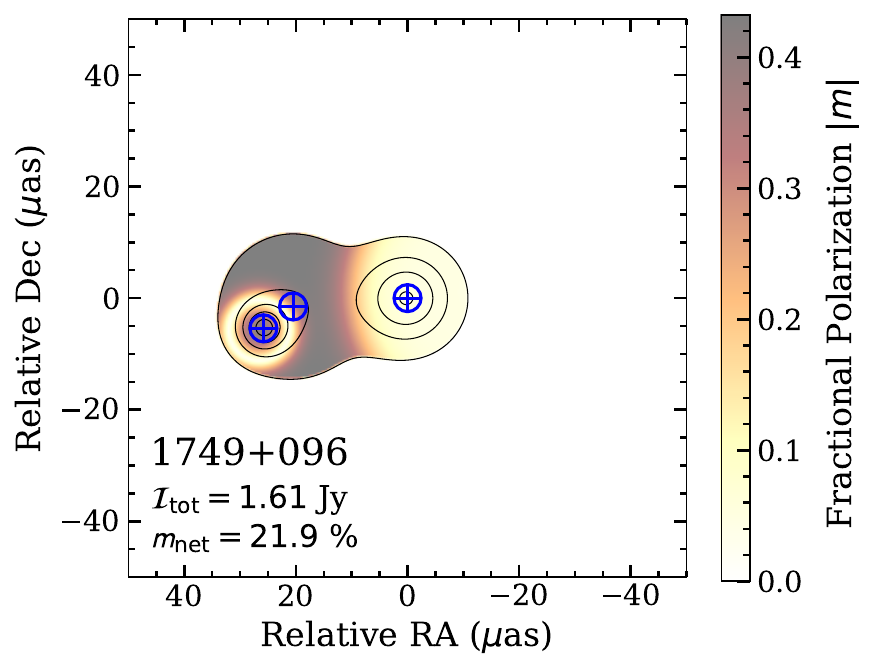}
    \hspace*{-0.1cm}\includegraphics[height=3.1cm,trim={1.1cm 0 1.1cm 0},clip]{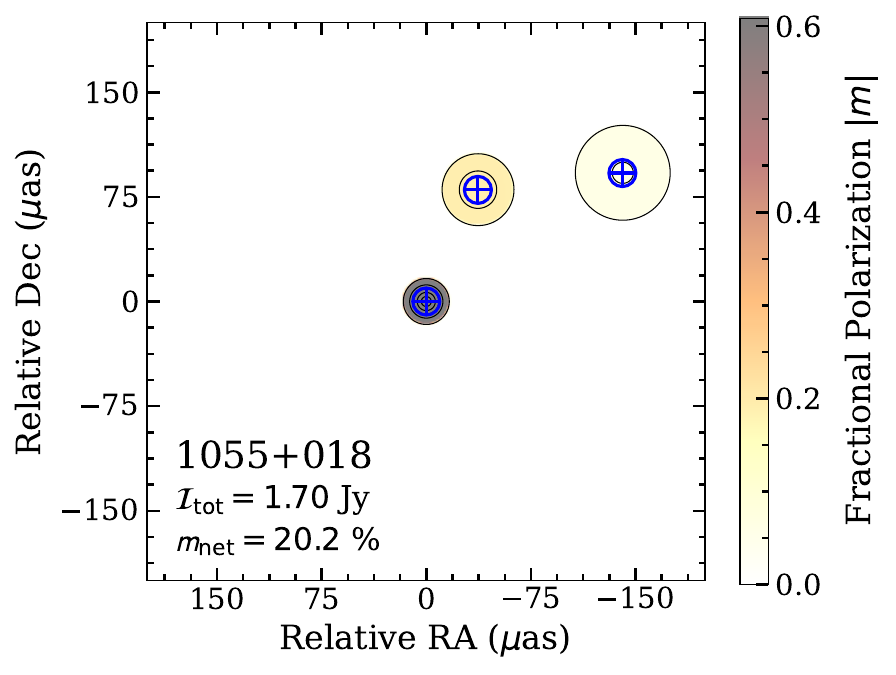}
    \includegraphics[height=3.1cm,trim={1.1cm 0 1.1cm 0},clip]{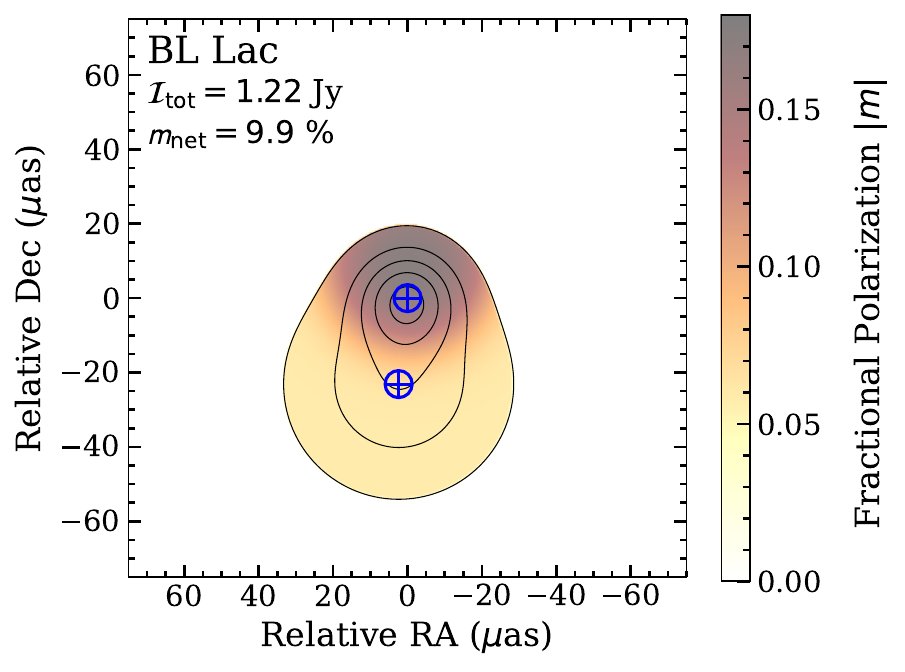}
    \includegraphics[height=3.1cm,trim={1.1cm 0 1.1cm 0},clip]{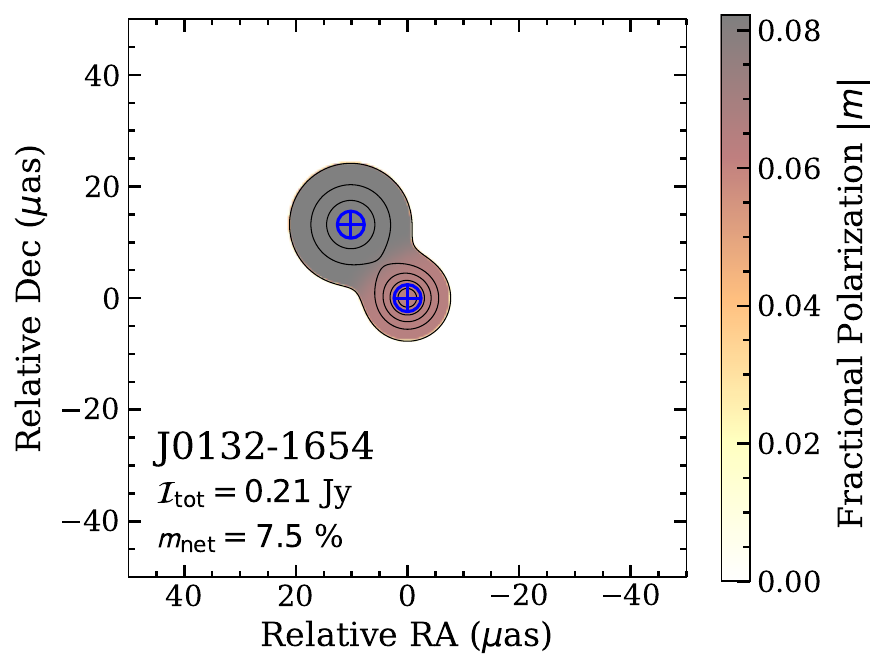}
    \includegraphics[height=3.1cm,trim={1.1cm 0 0cm 0},clip]{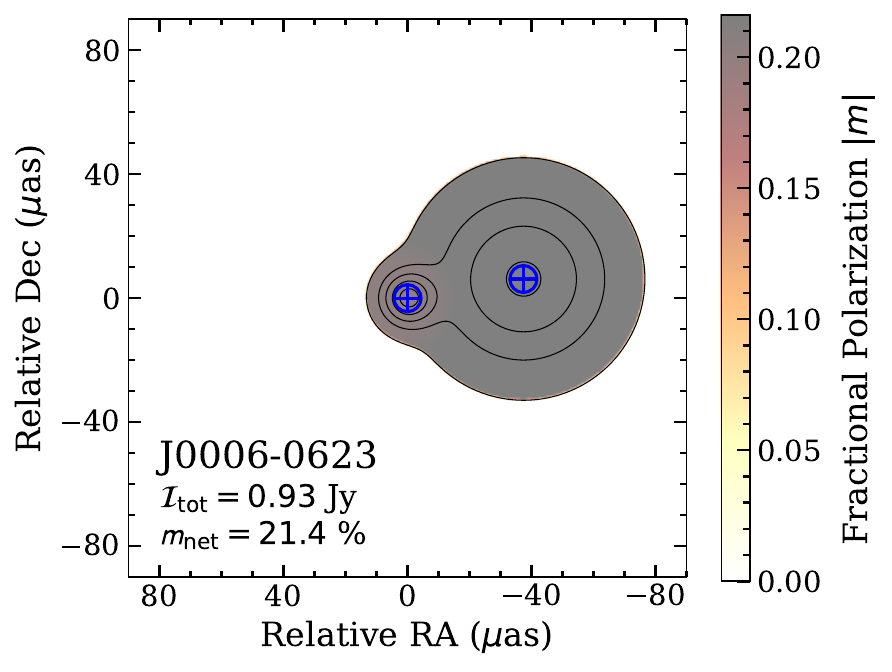}

 	\caption{Models of the EHT 2017 sources obtained through fitting of linearly polarized circular Gaussians. Contours represent 0.1, 0.3, 0.5, 0.7 and 0.9 of the peak total intensity brightness, same as in Fig.~\ref{fig:models_total}. Blue crosses: fitted positions of the Gaussians. Color map: fractional polarization. Total flux densities $I_{\rm tot}$ and total fractional polarization levels $m_{\rm net}$ for the reconstructed compact structures are annotated.} 
  \label{fig:models_info}
\end{figure*}

In Table~\ref{tab:models} we provide the results of geometric modeling for the seven sources observed during the EHT 2017 campaign that were not analysed in a 
separate publication. We considered circular Gaussian components with a constant fractional linear polarization. Our model-fitting procedure simultaneously 
minimizes errors on visibility amplitudes, closure phases, and fractional linear visibility polarization, see Section~\ref{subsec:modeling} and 
Fig.~\ref{fig:models_total}. We did not attempt to model circular polarization, given the very sparse sampling and low theoretically expected signal--to--noise ratio 
for the circular polarization. Furthermore, we did not generally fit for the amplitude gains, as they are very poorly constrained through interferometric closure 
quantities given such poor coverage. Instead, we incorporated $\sim\!10\%$ gains uncertainty into the error budget for the visibility amplitudes. While for the 
EHT observations with sufficient coverage and ALMA participation we performed the calibration of the polarization leakage and the absolute electric vector position angle, this is not possible 
for the seven sources discussed here. Hence, in Table~\ref{tab:models} we only provide the absolute fractional polarization of each component, without the 
polarimetric position angle information. For the five sources modeled with more than one circular Gaussian we show the corresponding maps of the fractional 
polarization in Fig.~\ref{fig:models_info}. In some cases (1749+096, 1055+018) we found extreme values of fractional polarization in the compact region. While 
the exact values are likely suffering from systematic biases related to sparse coverage and to a lack of the full polarization leakage calibration, the presence of high fractional polarization \rev{somewhere in the compact region} appears to be a robust 
result \rev{following our detections of high polarized correlated flux density on long baselines. While the wide field of view observations of AGN jets indicate the increase of the fractional polarization with the observing frequency \citep{Agudo2014}, some of the values that we estimate strongly exceed theoretical expectations for the optically thick emission from a VLBI core. This is puzzling and may indicate issues with the reconstructed morphology of the polarized emission or a reduction of the optical depth at 230\,GHz. } 
The systematic uncertainties are difficult to quantify reliably, depending not only on S/N and $(u,v)$-coverage but also non-trivially on the uncertain underlying source structure. Hence, we refrain from reporting untrustworthy uncertainties in Table~\ref{tab:models}. A conservative upper limit on the core brightness temperature uncertainty is a~factor of two difference between the measurement and the true value.

\begin{table*}[h!]
\caption{Circular Gaussian models for the EHT 2017 sources}
\vspace{-0.5cm}
\begin{center}
\begin{tabular}{llllllll}
\hline
\hline
Source &  Component & Flux (mJy) & FWHM ($\upmu$as) & $T_{\rm b} $ ($10^{10}$ K) & Distance ($\upmu$as) & PA (deg) & $p$ (\%)  \\ \hline

 \hline
1749+096   & Core & 196 & 6.6 & 13.9 & 0 & \quad-- & 65   \\ 

 & Jet 1 & 744 & 20.4 & 5.5 & 6.6 & --53.6 & 43 \\
 & Jet 2 & 674 & 12.9 & 12.6 & 26.4 & --78.2 & 4.6 \\

1055+018  & Core &  315 & 18.0 & 4.3 & 0 &\quad -- & 61 \\ 

 & Jet 1 & 531 & 34.8 & 1.9 & 88.1 & --24.8 & 19 \\
 & Jet 2 & 850 & 52.5 & 1.4 & 168.0 & --56.8 & 6.0 \\
 & Large scale & 1108 & -- & -- & -- &\quad -- & 3.6 \\

BL\,Lac    &  Core & 369 & 21.1 & 2.1 & 0 &\quad -- & 19   \\ 
 & Jet &  848 & 40.9 & 1.3 & 23.2 & $+$174.1  & 5.9  \\

J0132--1654 & Core & 80 & 8.5 & 5.3 & 0 & \quad-- & 6.4      \\ 
 & Jet & 132 & 13.2 & 3.6 & 16.6 & $+$37.7 & 8.2 \\
J0006--0623 & Core &  91 & 13.9 & 1.5 & 0 &\quad -- &  20  \\ 
 & Jet & 834 & 46.3 & 1.2 & 37.9 &  --81.1 & 22 \\
3C\,454.3 & Core  & 9040 & 22.5 & 77.7 & 0 &\quad -- & 17 \\  
CTA\,102  & Core & 5400 & 17.1 & 88.2 & 0 &\quad -- & 12   \\ 

\hline  
\label{tab:models}
\end{tabular}
\end{center}
\vspace{-0.7cm}
\tablefoot{PA: position angle with respect to the putative core (brightest component), east of north; $p$: fractional polarization of the component.}
\end{table*}

\clearpage
\FloatBarrier

\section{Doppler factor correction and power law fitting}
\label{Appendix:Doppler}

Following Section~\ref{subsection:PLmethod}, we used a dedicated procedure to characterize the dependence of observables with frequency using a power law model, while avoiding biases introduced by differences among individual sources, such as Doppler factor, intrinsic power, and distance. This procedure is labeled as "individual sources" in Table~\ref{tab:FitResults}, showing the same power law indices as reported in Section~\ref{sec:EHT+}. This approach, while chosen as favorable, is not unique though. Importantly, the observational sensitivity itself introduces a bias limiting the spread of the observed flux density values. As long as this spread is subdominant with respect to the measured variation with frequency, a large sample in which uncorrelated biases of individual sources average out, may be studied with a single power law fit to all sources \citep{Nair2019}. In other words, this is equivalent to fitting directly to the multi-source cloud of points in Fig.~\ref{fig:FluxSizeTb}. We only account for the cosmological redshift correction, which is small and does not impact the fit significantly. The fitting results are shown in the column "cosmology only" in Table~\ref{tab:FitResults}, indicating a decent consistency with our default method. If we further attempt to correct for the Doppler factor using estimates aggregated in Table~\ref{tab:BHmasses}, we increase the spread of points in the data as we act against the observational sensitivity bias that homogenized the observed flux densities. In other words, the spread grows as the data set is inhomogeneous in terms of the intrinsic (and not observed) source power. Thus, the fit to a Doppler-corrected cloud of multi-source data becomes less consistent with the default method, see "Doppler correction" column in Table~\ref{tab:FitResults}. One may attempt to increase the homogeneity of the sample by restricting the analysis to quasars and BL~Lacs, without radio galaxies ("Doppler, no radio galaxies" column in Table~\ref{tab:FitResults}). While this selection once again brings the estimated slopes to agreement with the default method, we neglect potentially useful data points. We conclude that our method based on aggregating power law fits to individual sources, insensitive to source-specific constant scaling like cosmological redshift or frequency-independent Doppler factor, is the preferable fitting approach.

\begin{table}[!htb]
\caption{Slopes of the power law fits obtained with different methods of fitting to data shown in Fig.~\ref{fig:FluxSizeTb}}
\vspace{-0.5cm}
\begin{center}
\begin{tabular}{lcccc}
\hline
\hline
               quantity  &  cosmology only & Doppler correction & Doppler, no radio galaxies & individual sources   \\ \hline
flux density $S_{\nu}$     & $-0.43\pm0.13$  & $-0.01\pm0.09$     & $-0.48\pm0.11$  & $-0.43\pm0.13$  \\ 
core size $\theta$    & $-0.80\pm0.07$  & $-0.72\pm0.04$     & $-0.77\pm0.06$  & $-0.64\pm0.05$  \\ 
brightness temperature $T_{\rm b}$ & $-0.96\pm0.16$  & $-0.46\pm0.10$     & $-0.93\pm0.13$  & $-0.95\pm0.13$  \\ 
\hline  
\label{tab:FitResults}
\end{tabular}
\end{center}
\vspace{-0.7cm}
\tablefoot{
Cosmology only: fits to the whole ensemble, corrected only for cosmological redshift; Doppler correction: Doppler correction applied to the data following Table~\ref{tab:BHmasses}; Doppler, no radio galaxies: same as the previous column, but excluding radio galaxies; individual sources: Mean of fit results to individual sources with the procedure described in Section~\ref{subsection:PLmethod}, shown as power laws in Fig.~\ref{fig:FluxSizeTb}.
}
\end{table}

In Table~\ref{tab:BHmasses} we compiled the estimates of black hole masses and Doppler factors for the EHT+ sample of AGN sources. Both are subject to large systematic uncertainties. For sources without a black hole mass estimate we assumed $M_{\sbullet[1.35]} = 10^8\,M_{\odot}$. For the Doppler factor corrections, we used values obtained from variability measurements at 15\,GHz (\citealt{Hovatta2009,Hovatta2009E,Liodakis2017}, compiled in \citealt{Pushkarev2017}), filling the gaps with 43\,GHz measurements from kinematics of characteristic jet components \citep{Weaver2022}. These measurements may not be of high accuracy, as the multi-frequency light curve variability of radio flares is not necessarily representative of the underlying jet kinematics, and the choice of a "characteristic" jet component is rather arbitrary. For the remaining sources we assumed $\delta = 1$ for radio galaxies and $\delta = 10$ for quasars and BL Lacs. 

\begin{table}[!htb]
\caption{Estimates for black hole mass and Doppler factor in the EHT+ AGN sample}
\vspace{-0.5cm}
\begin{center}
\begin{tabular}{lllll|l|lllll}
\hline
\hline
Name &  $M_{\sbullet[1.35]}\ (10^8\,M_\odot)$ & Ref.  & $\delta$ & Ref. & & Name &  $M_{\sbullet[1.35]}\ (10^8\,M_\odot)$ & Ref.  & $\delta$ & Ref.  \\ \hline


3C\,279    & $2.47\pm0.26$\tablefootmark{a}            & 1 & 23.8  & 7  & &
M\,87      & $65.0\pm2.0$\tablefootmark{b}             & 2 & 1.0   & --  \\ %
OJ\,287    & $\sim$1.0\tablefootmark{c}                & 3 & 16.8  & 7  & &
Cen\,A     &  $0.55\pm0.30$\tablefootmark{d}           & 4 & 1.0   &  -- \\ %
J1924--2914 & 1                                       & --  & 10.0  & --  & &
NRAO\,530  & $\sim$3.0\tablefootmark{e}                & 5 & 10.6  & 7  \\ %
3C\,273    & $7.83\pm2.50$\tablefootmark{a}            & 1 & 16.8  & 7  & &
1749+096   & 1                                       &  -- & 11.9  & 7  \\ %
1055+018   & $7.90\pm0.36$\tablefootmark{a}            & 1 & 12.1  & 7 & &
BL\,Lac    & $\sim$1.7\tablefootmark{f}                & 6 & 7.2   & 7  \\ %
3C\,84     & $\sim$3.0\tablefootmark{f}                & 6 & 6.9   & 8  & &
3C\,454.3  & $12.17\pm2.38$\tablefootmark{a}           & 1 & 32.9  & 8  \\ 
CTA\,102   & $6.39\pm0.24$\tablefootmark{a}            & 1 & 15.5  & 8  & &
J0132--1654 & 1                                       & --  & 10.0  & --  \\ %
J0006--0623 & 1                                       & --  & 5.1   & 7  & &
NGC\,1052  & $\sim$1.55\tablefootmark{f}               & 6 & 0.3   & 7  \\ 
\hline  
\label{tab:BHmasses}
\end{tabular}
\end{center}
\vspace{-0.7cm}
\tablefoot{
\tablefoottext{a}{spectral line widths}
\tablefoottext{b}{direct observation}
\tablefoottext{c}{scaling relations}
\tablefoottext{d}{stellar dynamics}
\tablefoottext{e}{spectral fitting}
\tablefoottext{f}{stellar velocity dispersion}\\
Doppler factors are estimated from 15\,GHz variability (7 \& refs. therein) or 43\,GHz kinematics (8). If a source is not available in (7), the measurement is taken from (8). We assumed a Doppler factor $\delta=1$ for the remaining radio galaxies (M\,87 and Cen\,A), and $\delta=10$ for the remaining quasars (J1924--2914 and J0006--0623).
}
\tablebib{
(1) \cite{Torrealba2012,Torrealba2012RMx}; 
(2) \cite{M87p1}; 
(3) \cite{Komossa2023}; 
(4) \cite{Neumayer2010};
(5) \cite{Keck2019};
(6) \cite{Woo2002};
(7) \cite{Pushkarev2017};
(8) \cite{Weaver2022}
}
\end{table}

\vfill

\FloatBarrier

\section{The flux density bias in the EHT VLBI data}
\label{app:VLBIbias}

\begin{figure*}[h!]
   \centering
	\includegraphics[width=0.35\columnwidth]{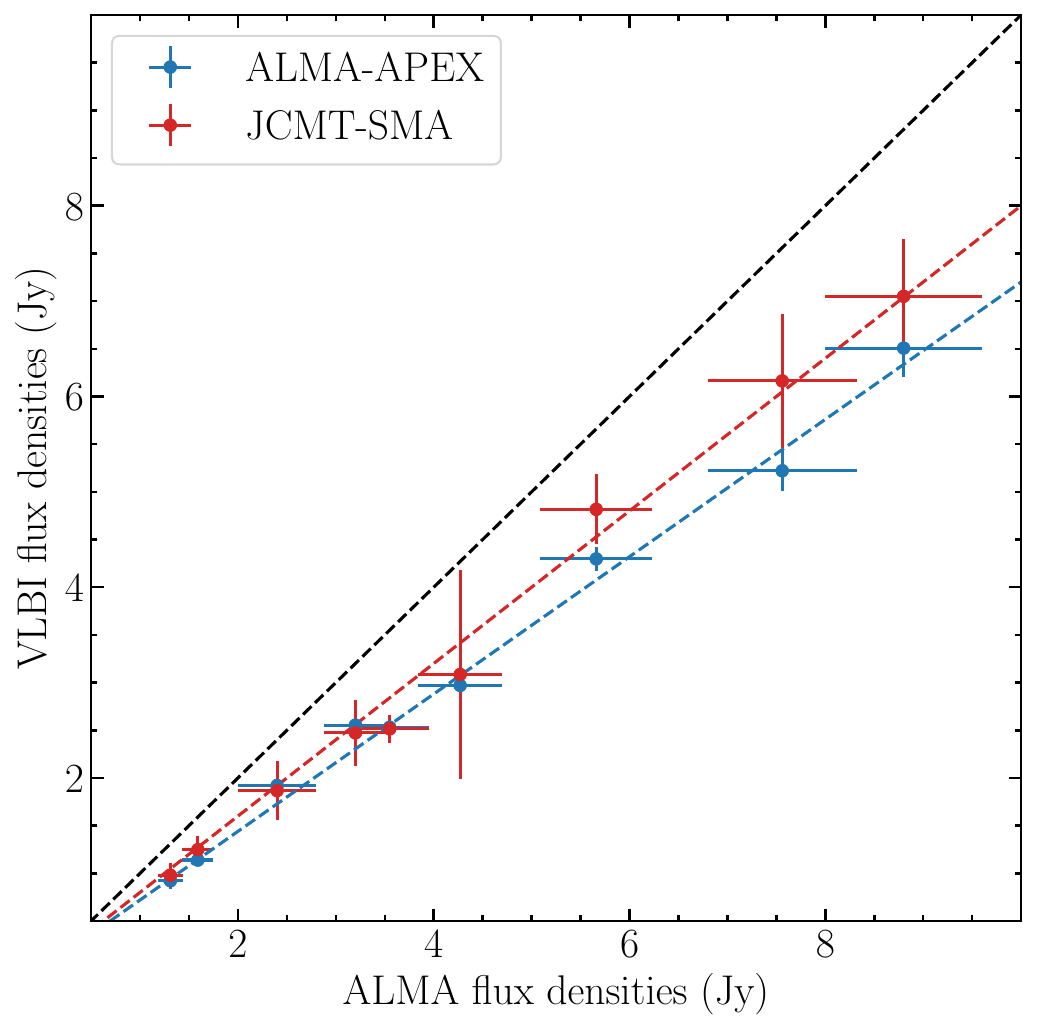}
	\includegraphics[width=0.35\columnwidth]{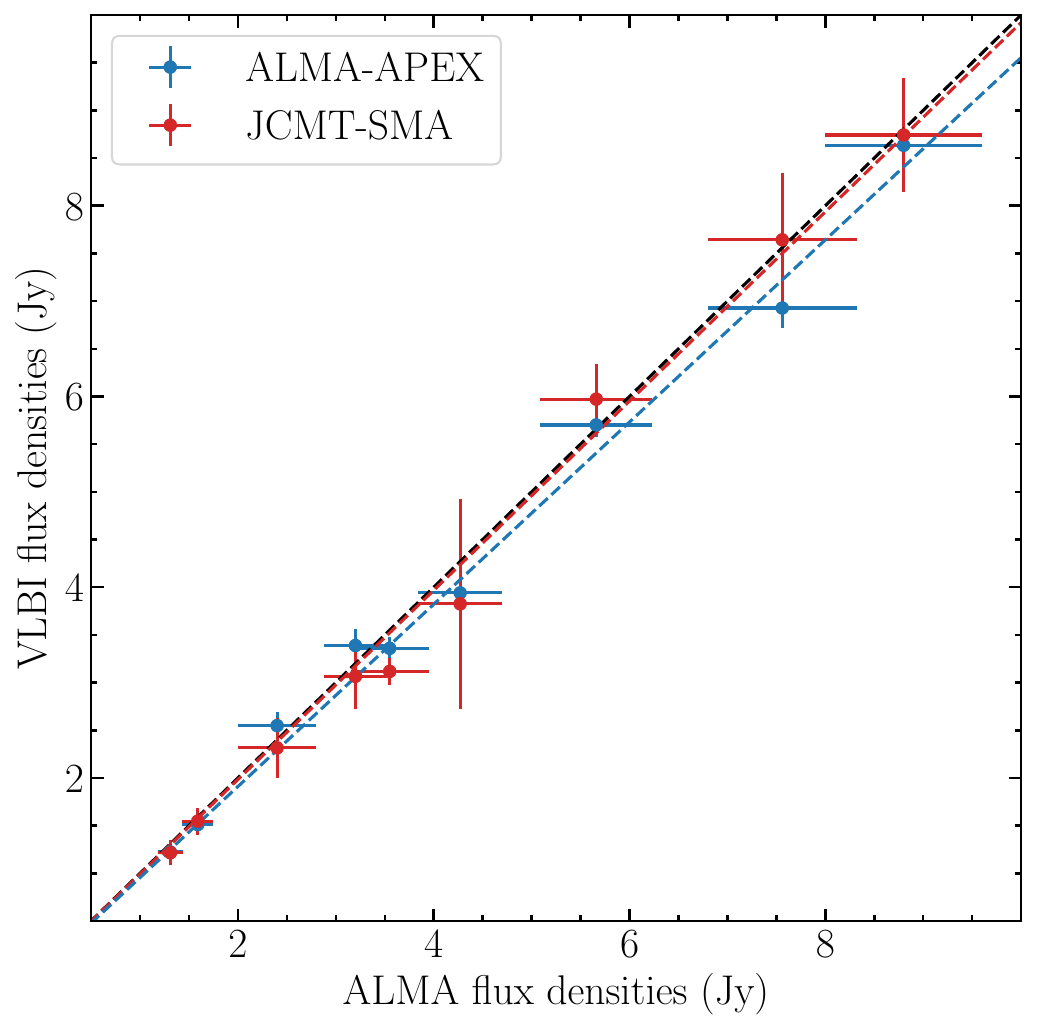}
    \caption{ALMA-only flux density measurements \citep{Goddi2021} vs intra-site VLBI flux densities obtained during the EHT campaign, as reported in Table~\ref{tab:sources}. The available data points correspond to observations of 3C279, 3C273, Centaurus~A, OJ287, 1055+018, J1924-2914, Sgr~A*, NRAO\,530, and M87. Left panel: the dashed lines scale ALMA-only flux densities down by 0.8 (red) and 0.72 (blue), to approximately match ALMA-APEX and JCMT-SMA measurements. Right panel: after correcting for 7\% ALMA VLBI losses and scaling all VLBI flux densities up by 25\%, we find consistency with ALMA-only data for all the sources.}
    \label{fig:VLBIbias}
\end{figure*}

We identify a systematic flux density deficit of 25\% between short intra-site VLBI baselines and the ALMA-only connected interferometric array measurements 
\citep{Goddi2021} after correcting for 7\% ALMA VLBI losses (G. Crew, private communication). This bias, illustrated in Fig.~\ref{fig:VLBIbias}, is generally 
avoided in the EHT calibration framework through the network calibration procedure \citep{M87p3,Blackburn2019,SgraP2}, scaling VLBI flux densities on intra-site baselines to ALMA measurements, whenever latter are available, through station-based amplitude gain calibration. 

Further investigations are necessary to pin 
down the exact cause of this effect. For the sources without ALMA-only data available, such as very sparse observations discussed in this paper, this bias is 
likely present, affecting the flux density measurements. We decided not to correct for this effect in case of sources modeled in this paper, as it is a) poorly characterized and it remains 
unclear if it affects all baselines in a uniform way, and b) because a bias of $\sim\!20\%$ is of little importance for our order of magnitude considerations, which are dominated by other systematic uncertainties.

\clearpage
\FloatBarrier


\section{Acknowledgements}
\label{app:acknowledgements}
\enlargethispage{\baselineskip}
\vspace{-0.2cm}
{\small 
We thank M. Sikora, K. Nalewajko, and A. Plavin for comments. J. Röder received financial support for this research from the International Max Planck Research School (IMPRS) for Astronomy and Astrophysics at the Universities of Bonn and Cologne. This research has made use of the SIMBAD database, operated at CDS, Strasbourg, France. This study makes use of VLBA data from the VLBA-BU Blazar Monitoring Program (BEAM-ME and VLBA-BU-BLAZAR, funded by NASA through the Fermi Guest Investigator Program). 

The Event Horizon Telescope Collaboration thanks the following
organizations and programs: the Academia Sinica; the Academy
of Finland (projects 274477, 284495, 312496, 315721); the Agencia Nacional de Investigaci\'{o}n 
y Desarrollo (ANID), Chile via NCN$19\_058$ (TITANs), Fondecyt 1221421  and BASAL FB210003; the Alexander
von Humboldt Stiftung; an Alfred P. Sloan Research Fellowship;
Allegro, the European ALMA Regional Centre node in the Netherlands, the NL astronomy
research network NOVA and the astronomy institutes of the University of Amsterdam, Leiden University, and Radboud University;
the ALMA North America Development Fund; the Astrophysics and High Energy Physics programme by MCIN (with funding from European Union NextGenerationEU, PRTR-C17I1); the Black Hole Initiative, which is funded by grants from the John Templeton Foundation (60477, 61497, 62286) and the Gordon and Betty Moore Foundation (Grant GBMF-8273) - although the opinions expressed in this work are those of the author and do not necessarily reflect the views of these Foundations
; the Brinson Foundation; ``la Caixa'' Foundation (ID 100010434) through fellowship codes LCF/BQ/DI22/11940027 and LCF/BQ/DI22/11940030; 
Chandra DD7-18089X and TM6-17006X; the China Scholarship
Council; the China Postdoctoral Science Foundation fellowships (2020M671266, 2022M712084); Conicyt through Fondecyt Postdoctorado (project 3220195); 
Consejo Nacional de Humanidades, Ciencia y Tecnología (CONAHCYT, Mexico, projects U0004-246083, U0004-259839, F0003-272050, M0037-279006, F0003-281692, 104497, 275201, 263356, 257435, CBF2023-2024-1102); the Colfuturo Scholarship; 
the Consejería de Economía, Conocimiento, 
Empresas y Universidad 
of the Junta de Andalucía (grant P18-FR-1769), the Consejo Superior de Investigaciones 
Científicas (grant 2019AEP112);
the Delaney Family via the Delaney Family John A.
Wheeler Chair at Perimeter Institute; Dirección General de Asuntos del Personal Académico-Universidad Nacional Autónoma de México (DGAPA-UNAM, projects IN112820 and IN108324); the Dutch Research Council (NWO) for the VICI award (grant 639.043.513), the grant OCENW.KLEIN.113, and the Dutch Black Hole Consortium (with project No. NWA 1292.19.202) of the research programme the National Science Agenda; the Dutch National Supercomputers, Cartesius and Snellius  
(NWO grant 2021.013); 
the EACOA Fellowship awarded by the East Asia Core
Observatories Association, which consists of the Academia Sinica Institute of Astronomy and
Astrophysics, the National Astronomical Observatory of Japan, Center for Astronomical Mega-Science,
Chinese Academy of Sciences, and the Korea Astronomy and Space Science Institute; 
the European Research Council (ERC) Synergy Grant ``BlackHoleCam: Imaging the Event Horizon of Black Holes" (grant 610058) and Synergy Grant "BlackHolistic: Colour Movies of Black Holes:
Understanding Black Hole Astrophysics from the Event Horizon to Galactic Scales" (grant 10107164); 
the European Union Horizon 2020
research and innovation programme under grant agreements
RadioNet (No. 730562),  
M2FINDERS (No. 101018682) and FunFiCO (No. 777740); the Horizon ERC Grants 2021 programme under grant agreement No. 101040021; the European Research Council for advanced grant `JETSET: Launching, propagation and 
emission of relativistic jets from binary mergers and across mass scales' (grant No. 884631); the European Horizon Europe staff exchange (SE) programme HORIZON-MSCA-2021-SE-01 grant NewFunFiCO (No. 10108625);
the FAPESP (Funda\c{c}\~ao de Amparo \'a Pesquisa do Estado de S\~ao Paulo) under grant 2021/01183-8; the Fondo CAS-ANID folio CAS220010; the Generalitat
Valenciana (grants APOSTD/2018/177 and  ASFAE/2022/018) and GenT Program (project CIDEGENT/2018/021); the Gordon and Betty Moore Foundation (GBMF-3561, GBMF-5278, GBMF-10423); the Institute for Advanced Study; the ICSC – Centro Nazionale di Ricerca in High Performance Computing, Big Data and Quantum Computing, funded by European Union – NextGenerationEU; the Istituto Nazionale di Fisica
Nucleare (INFN) sezione di Napoli, iniziative specifiche
TEONGRAV; 
the International Max Planck Research
School for Astronomy and Astrophysics at the
Universities of Bonn and Cologne; the Italian Ministry of University and Research (MUR)– Project CUP F53D23001260001, funded by the European Union – NextGenerationEU; 
DFG research grant ``Jet physics on horizon scales and beyond'' (grant No. 443220636);
Joint Columbia/Flatiron Postdoctoral Fellowship (research at the Flatiron Institute is supported by the Simons Foundation); 
the Japan Ministry of Education, Culture, Sports, Science and Technology (MEXT; grant JPMXP1020200109); 
the Japan Society for the Promotion of Science (JSPS) Grant-in-Aid for JSPS
Research Fellowship (JP17J08829); the Joint Institute for Computational Fundamental Science, Japan; the Key Research
Program of Frontier Sciences, Chinese Academy of
Sciences (CAS, grants QYZDJ-SSW-SLH057, QYZDJSSW-SYS008, ZDBS-LY-SLH011); 
the Leverhulme Trust Early Career Research
Fellowship; the Max-Planck-Gesellschaft (MPG);
the Max Planck Partner Group of the MPG and the
CAS; the MEXT/JSPS KAKENHI (grants 18KK0090, JP21H01137,
JP18H03721, JP18K13594, 18K03709, JP19K14761, 18H01245, 25120007, 23K03453); the MICINN Research Projects PID2019-108995GB-C22, PID2022-140888NB-C22 ; 
the MIT International Science
and Technology Initiatives (MISTI) Funds; 
the Ministry of Science and Technology (MOST) of Taiwan (103-2119-M-001-010-MY2, 105-2112-M-001-025-MY3, 105-2119-M-001-042, 106-2112-M-001-011, 106-2119-M-001-013, 106-2119-M-001-027, 106-2923-M-001-005, 107-2119-M-001-017, 107-2119-M-001-020, 107-2119-M-001-041, 107-2119-M-110-005, 107-2923-M-001-009, 108-2112-M-001-048, 108-2112-M-001-051, 108-2923-M-001-002, 109-2112-M-001-025, 109-2124-M-001-005, 109-2923-M-001-001, 110-2112-M-003-007-MY2, 110-2112-M-001-033, 110-2124-M-001-007, and 110-2923-M-001-001);
the Ministry of Education (MoE) of Taiwan Yushan Young Scholar Program;
the Physics Division, National Center for Theoretical Sciences of Taiwan;
the National Aeronautics and
Space Administration (NASA, Fermi Guest Investigator
grant 80NSSC20K1567, NASA Astrophysics Theory Program grant 80NSSC20K0527, NASA NuSTAR award 
80NSSC20K0645); 
NASA Hubble Fellowship 
grants HST-HF2-51431.001-A, HST-HF2-51482.001-A, HST-HF2-51539.001-A awarded 
by the Space Telescope Science Institute, which is operated by the Association of Universities for 
Research in Astronomy, Inc., for NASA, under contract NAS5-26555; 
the National Institute of Natural Sciences (NINS) of Japan; the National
Key Research and Development Program of China
(grant 2016YFA0400704, 2017YFA0402703, 2016YFA0400702);  the National Science and Technology Council (NSTC, grants NSTC 111-2112-M-001 -041, NSTC 111-2124-M-001-005, NSTC 112-2124-M-001-014); the US National
Science Foundation (NSF, grants AST-0096454,
AST-0352953, AST-0521233, AST-0705062, AST-0905844, AST-0922984, AST-1126433, OIA-1126433, AST-1140030,
DGE-1144085, AST-1207704, AST-1207730, AST-1207752, MRI-1228509, OPP-1248097, AST-1310896, AST-1440254, 
AST-1555365, AST-1614868, AST-1615796, AST-1715061, AST-1716327,  AST-1726637,
OISE-1743747, AST-1743747, AST-1816420, AST-1952099, AST-1935980, AST-2034306, AST-2205908, AST-2307887); 
NSF Astronomy and Astrophysics Postdoctoral Fellowship (AST-1903847); 
the Natural Science Foundation of China (grants 11650110427, 10625314, 11721303, 11725312, 11873028, 11933007, 11991052, 11991053, 12192220, 12192223, 12273022, 12325302, 12303021); 
the Natural Sciences and Engineering Research Council of
Canada (NSERC, including a Discovery Grant and
the NSERC Alexander Graham Bell Canada Graduate
Scholarships-Doctoral Program); the National Youth
Thousand Talents Program of China; the National Research
Foundation of Korea (the Global PhD Fellowship
Grant: grants NRF-2015H1A2A1033752, the Korea Research Fellowship Program:
NRF-2015H1D3A1066561, Brain Pool Program: 2019H1D3A1A01102564, 
Basic Research Support Grant 2019R1F1A1059721, 2021R1A6A3A01086420, 2022R1C1C1005255, 2022R1F1A1075115); the National Research Foundation of Korea (NRF) grant funded by the Korean government (MSIT) (RS-2024-00449206);
Netherlands Research School for Astronomy (NOVA) Virtual Institute of Accretion (VIA) postdoctoral fellowships; NOIRLab, which is managed by the Association of Universities for Research in Astronomy (AURA) under a cooperative agreement with the National Science Foundation;  
Onsala Space Observatory (OSO) national infrastructure, for the provisioning
of its facilities/observational support (OSO receives
funding through the Swedish Research Council under
grant 2017-00648);  

\pagebreak
\enlargethispage{\baselineskip}

\noindent
the Perimeter Institute for Theoretical
Physics (research at Perimeter Institute is supported
by the Government of Canada through the Department
of Innovation, Science and Economic Development
and by the Province of Ontario through the
Ministry of Research, Innovation and Science);  the Portuguese Foundation for Science and Technology (FCT) grants (Individual CEEC program - 5th edition, \url{https://doi.org/10.54499/UIDB/04106/2020}, \url{https://doi.org/10.54499/UIDP/04106/2020}, PTDC/FIS-AST/3041/2020, CERN/FIS-PAR/0024/2021, 2022.04560.PTDC); the Princeton Gravity Initiative; the Spanish Ministerio de Ciencia e Innovaci\'{o}n (grants PGC2018-098915-B-C21, AYA2016-80889-P,
PID2019-108995GB-C21, PID2020-117404GB-C21); 
the University of Pretoria for financial aid in the provision of the new 
Cluster Server nodes and SuperMicro (USA) for a SEEDING GRANT approved toward these 
nodes in 2020; the Shanghai Municipality orientation program of basic research for international scientists (grant no. 22JC1410600);
the Shanghai Pilot Program for Basic Research, Chinese Academy of Science, 
Shanghai Branch (JCYJ-SHFY-2021-013); the Simons Foundation (grant 00001470); 
the State Agency for Research of the Spanish MCIU through
the ``Center of Excellence Severo Ochoa'' award for
the Instituto de Astrof\'{i}sica de Andaluc\'{i}a (SEV-2017-
0709); the Spanish Ministry for Science and Innovation grant CEX2021-001131-S funded by MCIN/AEI/10.13039/501100011033; the Spinoza Prize SPI 78-409; the South African Research Chairs Initiative, through the 
South African Radio Astronomy Observatory (SARAO, grant ID 77948),  which is a facility of the National 
Research Foundation (NRF), an agency of the Department of Science and Innovation (DSI) of South Africa; 
the Toray Science Foundation; the Swedish Research Council (VR); the UK Science and Technology Facilities Council (grant no. ST/X508329/1); 
the US Department
of Energy (USDOE) through the Los Alamos National
Laboratory (operated by Triad National Security,
LLC, for the National Nuclear Security Administration
of the USDOE, contract 89233218CNA000001); and the YCAA Prize Postdoctoral Fellowship.

We thank
the staff at the participating observatories, correlation
centers, and institutions for their enthusiastic support.
This paper makes use of the following ALMA data:
ADS/JAO.ALMA\#2016.1.00413.V, ADS/JAO.ALMA\#2016.1.01114.V, 
ADS/JAO.ALMA\#2016.1.01116.V, ADS/JAO.ALMA\#2016.1.01154.V,
ADS/JAO.ALMA\#2016.1.01176.V, ADS/JAO.ALMA\#2016.1.01198.V, 
ADS/JAO.ALMA\#2016.1.01216.V, ADS/JAO.ALMA\#2016.1.01290.V, 
ADS/JAO.ALMA\#2016.1.01404.V, ADS/JAO.ALMA\#2017.1.00841.V, 
ADS/JAO.ALMA\#2013.1.01022.S, ADS/JAO.ALMA\#2015.1.01170.S, 
ADS/JAO.ALMA\#2016.1.00415.S, ADS/JAO.ALMA\#2017.1.00608.S. 
ALMA is a partnership
of the European Southern Observatory (ESO;
Europe, representing its member states), NSF, and
National Institutes of Natural Sciences of Japan, together
with National Research Council (Canada), Ministry
of Science and Technology (MOST; Taiwan),
Academia Sinica Institute of Astronomy and Astrophysics
(ASIAA; Taiwan), and Korea Astronomy and
Space Science Institute (KASI; Republic of Korea), in
cooperation with the Republic of Chile. The Joint
ALMA Observatory is operated by ESO, Associated
Universities, Inc. (AUI)/NRAO, and the National Astronomical
Observatory of Japan (NAOJ). The NRAO
is a facility of the NSF operated under cooperative agreement
by AUI.

This research used data obtained with the Global Millimeter VLBI Array (GMVA). The
GMVA consists of telescopes operated by the MPIfR, IRAM, Onsala, Metsahovi, Yebes, the Korean VLBI Network, the Green Bank Observatory and the Very Long Baseline Array (VLBA). The VLBA and the GBT are facilities of the National Science Foundation operated under cooperative agreement by Associated Universities, Inc. GMVA data were correlated at the Max Planck Institute for Radio Astronomy (Bonn, Germany). This research used resources of the Oak Ridge Leadership Computing Facility at the Oak Ridge National
Laboratory, which is supported by the Office of Science of the U.S. Department of Energy under contract
No. DE-AC05-00OR22725; the ASTROVIVES FEDER infrastructure, with project code IDIFEDER-2021-086; the computing cluster of Shanghai VLBI correlator supported by the Special Fund 
for Astronomy from the Ministry of Finance in China;  
We also thank the Center for Computational Astrophysics, National Astronomical Observatory of Japan. This work was supported by FAPESP (Fundacao de Amparo a Pesquisa do Estado de Sao Paulo) under grant 2021/01183-8. 

APEX is a collaboration between the
Max-Planck-Institut f{\"u}r Radioastronomie (Germany),
ESO, and the Onsala Space Observatory (Sweden). The
SMA is a joint project between the SAO and ASIAA
and is funded by the Smithsonian Institution and the
Academia Sinica. The JCMT is operated by the East
Asian Observatory on behalf of the NAOJ, ASIAA, and
KASI, as well as the Ministry of Finance of China, Chinese
Academy of Sciences, and the National Key Research and Development
Program (No. 2017YFA0402700) of China
and Natural Science Foundation of China grant 11873028.
Additional funding support for the JCMT is provided by the Science
and Technologies Facility Council (UK) and participating
universities in the UK and Canada. 
The LMT is a project operated by the Instituto Nacional
de Astr\'{o}fisica, \'{O}ptica, y Electr\'{o}nica (Mexico) and the
University of Massachusetts at Amherst (USA). The
IRAM 30-m telescope on Pico Veleta, Spain is operated
by IRAM and supported by CNRS (Centre National de
la Recherche Scientifique, France), MPG (Max-Planck-Gesellschaft, Germany), 
and IGN (Instituto Geogr\'{a}fico
Nacional, Spain). The SMT is operated by the Arizona
Radio Observatory, a part of the Steward Observatory
of the University of Arizona, with financial support of
operations from the State of Arizona and financial support
for instrumentation development from the NSF.
Support for SPT participation in the EHT is provided by the National Science Foundation through award OPP-1852617 
to the University of Chicago. Partial support is also 
provided by the Kavli Institute of Cosmological Physics at the University of Chicago. The SPT hydrogen maser was 
provided on loan from the GLT, courtesy of ASIAA.

This work used the
Extreme Science and Engineering Discovery Environment
(XSEDE), supported by NSF grant ACI-1548562,
and CyVerse, supported by NSF grants DBI-0735191,
DBI-1265383, and DBI-1743442. XSEDE Stampede2 resource
at TACC was allocated through TG-AST170024
and TG-AST080026N. XSEDE JetStream resource at
PTI and TACC was allocated through AST170028.
This research is part of the Frontera computing project at the Texas Advanced 
Computing Center through the Frontera Large-Scale Community Partnerships allocation
AST20023. Frontera is made possible by National Science Foundation award OAC-1818253.
This research was done using services provided by the OSG Consortium~\citep{osg07,osg09} supported by the National Science Foundation award Nos. 2030508 and 1836650. 
Additional work used ABACUS2.0, which is part of the eScience center at Southern Denmark University, and the Kultrun Astronomy Hybrid Cluster (projects Conicyt Programa de Astronomia Fondo Quimal QUIMAL170001, Conicyt PIA ACT172033, Fondecyt Iniciacion 11170268, Quimal 220002). 
Simulations were also performed on the SuperMUC cluster at the LRZ in Garching, 
on the LOEWE cluster in CSC in Frankfurt, on the HazelHen cluster at the HLRS in Stuttgart, 
and on the Pi2.0 and Siyuan Mark-I at Shanghai Jiao Tong University.
The computer resources of the Finnish IT Center for Science (CSC) and the Finnish Computing 
Competence Infrastructure (FCCI) project are acknowledged. This
research was enabled in part by support provided
by Compute Ontario (http://computeontario.ca), Calcul
Quebec (http://www.calculquebec.ca), and Compute
Canada (http://www.computecanada.ca). 

The EHTC has
received generous donations of FPGA chips from Xilinx
Inc., under the Xilinx University Program. The EHTC
has benefited from technology shared under open-source
license by the Collaboration for Astronomy Signal Processing
and Electronics Research (CASPER). The EHT
project is grateful to T4Science and Microsemi for their
assistance with hydrogen masers. This research has
made use of NASA's Astrophysics Data System. We
gratefully acknowledge the support provided by the extended
staff of the ALMA, from the inception of
the ALMA Phasing Project through the observational
campaigns of 2017 and 2018. We would like to thank
A. Deller and W. Brisken for EHT-specific support with
the use of DiFX. We thank Martin Shepherd for the addition of extra features in the Difmap software 
that were used for the CLEAN imaging results presented in this paper.
We acknowledge the significance that
Maunakea, where the SMA and JCMT EHT stations
are located, has for the indigenous Hawaiian people.
}
\vfill


\end{document}